\documentclass[showpacs,preprintnumbers,amsmath,amssymb]{revtex4}

\usepackage{amssymb}
\usepackage{amsmath}
\usepackage{mathrsfs}
\usepackage{graphics}
\usepackage{epsfig}
\usepackage{graphicx}
\usepackage{subfigure}
\usepackage{bm}
\usepackage{nicefrac}
\usepackage{feynmp}

\def\beq{\begin{equation}}
\def\eeq{\end{equation}}
\def\baq{\begin{eqnarray}}
\def\eaq{\end{eqnarray}}

\def\be{\begin{equation}}
\def\ee{\end{equation}}
\def\bea{\begin{eqnarray}}
\def\eea{\end{eqnarray}}

\def\rs{\rho_{\ast}}
\def\UV{\text{UV}}

\def\ap{\alpha'}
\def\apsq{\alpha'^{2}}
\def\gs{g_{s}}
\def\ruv{\r_{\UV}}
\def\mp{M_{P}}

\def\eg{{\it e.\,g.}~}
\def\ie{{\it i.\,e.}~}

\def\mcA{{\mathcal A}}
\def\mcB{{\mathcal B}}
\def\mcC{{\mathcal C}}

\def\mcM{{\mathcal M}}
\def\mcN{{\mathcal N}}
\def\mcO{{\mathcal O}}

\def\mcU{{\mathcal U}}
\def\mcV{{\mathcal V}}

\def\k{{\bf k}}

\def\to{\rightarrow}

\def\to{\rightarrow}

\def\a{\alpha}

\def\e{\epsilon}           

\def\k{\kappa}                    

  \def\w{\omega}
\def\r{\rho}                                     
\def\s{\sigma}                                   

\def\6{\partial}

\newcommand{\bright}{\begin{flushright}}
\newcommand{\eright}{\end{flushright}}
\newcommand{\bminip}{\begin{minipage}}
\newcommand{\eminip}{\end{minipage}}
\newcommand{\bcent}{\begin{center}}
\newcommand{\ecent}{\end{center}}

\begin{document}

\unitlength = 1mm

\begin{flushleft}
MPP-2012-138 \hfill October 26, 2012 \hfill LMU-ASC 74/12
\end{flushleft}

\title{Slow-walking inflation}

\author{Johanna Erdmenger ${}^{(1)}$, Sebastian Halter ${}^{(1,2)}$, Carlos N\'u\~nez ${}^{(3)}$, Gianmassimo Tasinato${}^{(4)}$ \footnote{jke@mppmu.mpg.de, s.halter@physik.uni-muenchen.de, c.nunez@swansea.ac.uk, gianmassimo.tasinato@port.ac.uk} \\
}

\affiliation{%
\hskip1cm
\\
${}^{(1)}$ \it{Max-Planck-Institut f\"ur Physik, F\"ohringer Ring 6,
  80805 M\"unchen, Germany.} \\
${}^{(2)}$ \it{Arnold-Sommerfeld-Center for Theoretical Physics,
\\ Fakult\"at f\"ur Physik, 
Ludwig-Maximilians-Universit\"at M\"unchen,\\ 
Theresienstra{\ss}e 37, 80333 M\"unchen, Germany.} \\
${}^{(3)}$ \it{Department of Physics,\\ University of Swansea, Singleton 
Park\\
Swansea SA2 8PP,  United Kingdom.} \\
${}^{(4)}$ Institute of Cosmology $\&$ Gravitation, University of Portsmouth,
 Portsmouth, PO1 3FX, United Kingdom.\\ }
 
\begin{abstract}
We propose a new model of slow-roll inflation in string cosmology, based on warped throat supergravity solutions displaying `walking' dynamics, \ie the coupling constant of the dual gauge theory slowly varies over a range of energy scales. The features of the throat geometry are sourced by a rich field content, given by  the dilaton and RR and NS fluxes. By considering the motion of a D3-brane probe in this geometry, we are able to analytically calculate the brane potential in a physically interesting regime. This potential has an inflection point: in its proximity we realize a model of inflation lasting sixty e-foldings, and whose robust predictions are in agreement with current observations. We are also able to interpret some of the most interesting aspects  of this scenario in terms of  the properties of the QFT dual theory.
\end{abstract}

\maketitle

\smallskip





\section{Introduction}

The inflationary paradigm provides convincing solutions to the basic problems of standard big bang cosmology. Moreover, it offers a testable mechanism for the generation of primordial cosmological perturbations that seed the formation of large scale structures and are imprinted on the cosmological microwave background radiation \cite{Lyth:2009zz}. So far, the simplest models of inflation are in agreement with observations; they are based on a single scalar field, the inflaton, which slowly rolls down a flat potential for a period sufficiently long to achieve sixty $e$-folds of inflation. 

\smallskip

While at first sight it seems easy to build inflationary scenarios with the desired properties (and indeed there are plenty of proposals), a theoretical challenge is to embed consistent models of inflation in a fundamental theory, capable to combine cosmological inflation with particle physics models at short distances. An option is to embed inflation in string theory, using as inflaton candidate one of the many light moduli available in string set-ups. See \cite{McAllister:2007bg} for recent nice reviews that also refer to the relevant literature.

\smallskip

Much work has been done in particular for embedding models of
inflation into warped flux compactifications of type IIB string theory. This approach is
referred to as \emph{warped D-brane inflation} \cite{Kachru:2003sx}, in which inflation is
realized by the motion of a probe D3-brane along the radial direction
of a \emph{strongly-warped throat} region inside a flux compactification. In this type of geometries, fluxes \cite{Giddings:2001yu} and non-perturbative
effects, such as gaugino condensation on wrapped D7-branes \cite{Kachru:2003aw}, are turned
on to stabilize light moduli that do not drive inflation. This
approach has been refined significantly over past years, using the
deep understanding of warped conifold geometries recently acquired,
motivated by the AdS/CFT correspondence. 

In this work, we  continue to build along this line of research, embedding a slow-roll model of inflation 
in a warped throat geometry with various background fields and fluxes
turned on. We will focus on a warped throat dual to a gauge theory
that exhibits  `walking behavior' \cite{Nunez:2008wi,Elander:2009pk}, \ie which has a coupling constant characterized by an RG flow that runs slowly within a range of energy scales. From a geometrical point of view, this energy range corresponds to  a region of transition between a geometry sourced by a stack of D5-branes in the deep
infrared (IR), and a space-time that approximates the Klebanov-Strassler (KS) throat \cite{Klebanov:2000hb} 
in the ultraviolet (UV). This region is controlled by fluxes that
acquire non-trivial profiles which are tunable to a certain extent.  
The rich content of background fields turned on in the geometry ,
which is necessary to generate multiple scale dynamics in the QFT
dual, offers an attractive arena for embedding models of warped
D-brane inflation. This is the scope of this work, in which we show
that a successful inflationary model can be embedded in this
geometry using a probe D3-brane. Fluxes and warping generate a force
which acts on the probe D3-brane, thereby inducing inflation with no
need to include additional sources such as anti-D3-branes: in this
sense, our scenario can be seen as a generalization of the
inflationary model built on the baryonic branch of the
Klebanov-Strassler  throat \cite{Dymarsky:2005xt}. In our set-up,
slow-roll inflation naturally occurs precisely in the region in which
the walking dynamics is manifest in the dual field theory; hence the
name \emph{slow-walking inflation}. Remarkably, in the regime we are
focussing on, we find a fully analytical form for the probe brane potential,
which enables us to  clearly appreciate the roles of the geometrical
parameters that govern this potential. From the perspective of the
dual field theory, it is possible to identify which operators are
responsible for generating the main features of the  potential,
leading to a QFT interpretation of some aspects of our scenario. In
particular, the VEVs of two operators in the QFT dual start to
dominate in the walking regime, and combine in such a way to generate
the flat potential for the D3-brane. The large number of parameters
characterizing the geometry provide enough freedom to tune the
D-brane potential, allowing to comfortably accommodate
the requirements of producing sixty $e$-folds of slow-roll inflation,
and to match the observed amplitude and scale-dependence for the power spectrum of scalar fluctuations. Interestingly, inflation occurs naturally around an \emph{inflection point} of the brane potential, which can be related to the non-trivial profile of the dilaton along the radial direction. Within our scenario, we obtain robust predictions for the properties of the inflationary process:  the slow-roll $\epsilon$-parameter is much smaller than the $\eta$-parameter; the $\eta$-parameter is negative, and its absolute value is proportional to the inverse of the number of $e$-folds. The resulting value for the scalar spectral index is in good agreement with observations by WMAP \cite{Komatsu:2010fb}.
\smallskip

Building a model of inflation in the warped throat geometry, as we
achieve here, is only the first step in the theoretical investigation
of our scenario. Next, it needs to be checked whether inflation survives after including corrections to the D-brane potential due to couplings between the D3-brane, the moduli stabilizing sectors and gravitational degrees of freedom as well as Kaluza-Klein modes. This is the so-called $\eta$-problem: typically, corrections to the D3-brane potential due to the aforementioned effects tend to make the $\eta$-parameter large, and spoil inflation.
 Originally, this problem was pointed out in \cite{Copeland:1994vg} when attempting to embed inflation in supergravity. It has been widely analyzed in recent years in the context of warped D-brane inflation \cite{Kachru:2003sx,Berg:2004sj,Baumann:2006cd,Burgess:2006cb,Baumann:2006th,Krause:2007jk,Chen:2009nk,Baumann:2008kq,Baumann:2009qx,Baumann:2010sx,Agarwal:2011wm,Dias:2012nf,McAllister:2012am}, when embedding a D3-brane in the AdS-like region of a Klebanov-Strassler throat. Our set-up is different from the KS warped throat D-brane scenario. Thus, a careful analysis of the $\eta$-problem in the present context is not straightforward so we will not pursue it here. Let us however point out that the large number of parameters involved and the various operators that can be switched on from the QFT perspective, leave the hope that dangerous contributions from compactification effects can be tuned away without qualitatively changing the features of our set-up.

\smallskip

The plan of the paper is the following. In Section~\ref{Sec-WalkingGeometry}, we carefully describe the geometry that constitutes the arena for our inflationary model, with emphasis on the parameters describing the warped throat (see in particular Section~\ref{Sec-PhysCharactGeometry}). In Section~\ref{Sec-ReviewD3Inflation}, we briefly review D3-brane inflation in a generic geometry of the above form. In Section~\ref{Sec-SlowWalkingInflation}, these results are then applied for the background of Section~\ref{Sec-WalkingGeometry}. We find (for a particular range of parameters) analytic expressions for the quantities for inflation ($\epsilon, \eta, N_e$ and the amplitude of the power spectrum), showing that they exhibit a scaling behavior with the parameters of the model. An explicit inflationary trajectory is also presented. Finally, Section~\ref{Sec-Discussion} summarizes our results and concludes pointing to future possible studies. Various appendices of technical nature are included. We note that other models of inflation based on embedding probe branes within gauge/gravity duality have recently been discussed in \cite{Chen:2010pd,Evans:2010tf,Evans:2012jx,Nastase:2011qz,Channuie:2011rq,Bezrukov:2011mv,Channuie:2012bv}.


\section{The walking geometry}
\label{Sec-WalkingGeometry}

In this section, we discuss the warped throat geometry of interest, emphasizing its geometrical and physical aspects that we will use later for building our model of inflation. The same class of geometries were considered in \cite{Nunez:2008wi,Elander:2009pk}, where they were instead used for building particle physics models of walking technicolor.

\subsection{Wrapped-D5 system}
\label{Sec-GeneratingWrappedD5}

In this section, we will consider
type IIB string theory in the Supergravity limit. The topology of the background space time will contain a four dimensional Minkowski space and a 
six-dimensional Calabi-Yau (CY) `internal' space related to the conifold. Due to this, the background space-time will preserve minimal ($\mcN= 1$ supersymmetry
in four dimensions.

Let us start then, by considering type IIB supergravity and  the geometry produced by stacking on top of each other $N_c$ D5-branes wrapping an $S^2$ inside a 6d Calabi-Yau (CY) cone \cite{Maldacena:2000yy,HoyosBadajoz:2008fw} --- \eg the conifold with topology $\mathbb{R} \times S^{2} \times S^{3}$. We truncate type IIB supergravity to include only the metric, the dilaton $\Phi$ and the RR three-form $F_3$, which we express in terms of the ($SU(2)$ left-invariant) one-forms $\tilde{\w}_i$, ($i=1,2,3$) --- see Appendix~\ref{App-BackgroundDetails} for more details. We then propose  an ansatz that assumes the functions appearing in the background to depend only on the radial coordinate $\r$. Setting $\ap \gs =1$, we write the background metric in Einstein frame as
\be
\label{Eq-WrappedD5Metric}
\begin{split}
  ds_{E}^2 = \; & e^{\Phi(\rho)/2} \Big[ dx_{1,3}^2 + e^{2k(\rho)}d\rho^2 + e^{2 q(\rho)} (d\theta^2 + \sin^2\theta d\phi^2) \\
  & + \frac{e^{2 {g}(\rho)}}{4} \left[(\tilde{\omega}_1+a(\rho)d\theta)^2 + (\tilde{\omega}_2-a(\rho)\sin\theta d\phi)^2\right]  + \frac{e^{2 k(\rho)}}{4} (\tilde{\omega}_3 + \cos\theta d\phi)^2\Big] \, .
\end{split}
\ee
As mentioned above, there are also a dilaton $\Phi$ and a RR-three form $F_{3}$ that complete the background. We write details of the full configuration in Appendix~\ref{App-BackgroundDetails}, which as we commented, will preserve minimal SUSY.

Any particular SUSY solution is determined by solving the system of BPS equations. These equations are in general non-linear and coupled. They can be rearranged in a convenient form by rewriting the functions appearing in Eq.~\eqref{Eq-WrappedD5Metric} in terms of a new set of functions (a `basis' that we call $P,Q,....$ defined in Eq.~\eqref{Eq-NewFunctions1Appendix} in Appendix~\ref{App-BackgroundDetails}), to make the system still non-linear but at least  decoupled.

After some lengthy  algebra (described in detail in \cite{HoyosBadajoz:2008fw} and summarized in Appendix~\ref{App-BackgroundDetails}), we find that the relevant functions for the purposes of this paper are given by
\be
\label{Eq-NewFunctions2}
Q(\rho) = N_c (2 \rho \coth (2\rho) -1) \, , \quad 2 e^{2k} = P' \, , \quad e^{4\Phi} = \frac{2\,e^{4\Phi_0} \,\sinh^2(2\rho)  }{(P^2-Q^2) P'} \, ,
\ee
with $\Phi_0$ an integration constant and the function $P(\r)$ satisfying the second order differential equation
\beq
\label{Eq-MasterEquation}
P'' + P' \left( \frac{P' + Q'}{P - Q} +\frac{P' - Q'}{P + Q} - 4 \coth(2\rho) \right) = 0 \, .
\eeq
We will refer to Eq.~\eqref{Eq-MasterEquation} as the \emph{master equation}. This is the \emph{only} equation that needs solving in order to generate the large classes of supersymmetric background solutions we are interested in. The equation for $P$ depends on the function $Q$ defined in Eq.~\eqref{Eq-NewFunctions2}, proportional to the number $N_c$ of wrapped branes. 

\subsection{Rotation: U-duality as a solution-generating technique}
\label{Sec-GeneratingRotationUDuality}

We now take the configuration of Eq.~\eqref{Eq-WrappedD5Metric} and apply to it a solution generating technique. This will produce a more interesting background in view of our applications (more on this in subsection~\ref{Sec-PhysCharactGeometry}). In \cite{Maldacena:2009mw}, the authors proposed a U-duality that takes a background of the form written in Eq.~\eqref{Eq-WrappedD5Metric} with certain particular solutions of Eq.~\eqref{Eq-MasterEquation} and maps them into another background where new fluxes are turned on. See also \cite{Gaillard:2010qg} for a different perspective on this technique.

The effect of this solution-generating technique (which we will refer to as \emph{`rotation'}) can be summarized by defining a basis of 1-forms, which in Einstein frame reads
\be
\label{Eq-Vielbein}
\begin{split}
  e^{xi} & = \hat{h}^{-\frac{1}{4}}e^{\frac{\Phi}{4}}dx_i \, , \quad e^{\r} = k_1^\frac12 \hat{h}^{\frac{1}{4}}e^{\frac{\Phi}{4} +k}d\r \, , \quad e^{3} = k_1^\frac12 \hat{h}^{\frac{1}{4}}\frac{e^{\frac{\Phi}{4} +k}}{2} (\tilde{\w}_3+\cos\theta d\varphi) \, , \\
  e^{\theta} & =  k_1^\frac12 \hat{h}^{\frac{1}{4}}e^{\frac{\Phi}{4} +q} d\theta \, , \quad e^{\varphi} = k_1^\frac12 \hat{h}^{\frac{1}{4}}e^{\frac{\Phi}{4} +q}\sin\theta d\varphi \, , \\
  e^{1} & =  k_1^\frac12 \hat{h}^{\frac{1}{4}}\frac{e^{\frac{\Phi}{4} +g}}{2}(\tilde{\w}_1+a d\theta ) \, , \quad e^{2} =  k_1^\frac12 \hat{h}^{\frac{1}{4}}\frac{e^{\frac{\Phi}{4} +g}}{2}(\tilde{\w}_2 -a\sin\theta d\varphi) \, ,
\end{split}
\ee
where $x_i$ are the four Minkowski directions (and the $\tilde{\w}_{i}$ are defined in Eq.~\eqref{Eq-SU2Inv1Forms}). We define the warp factor $\hat{h}$ appearing in the previous expression as
\be
\label{Eq-RotatedWarpFactor}
\hat{h} \equiv k_1^{-2} \left(1 - k_{2}^{2} \, e^{2\Phi} \right) \, ,
\ee
which depends on two constants, $k_1$ and $k_2$. These will be determined later on by appropriate physical requirements. The new configuration contains a metric $g_{M N}$, a dilaton $\Phi$, a RR three-form $F_3$ and a five-form $F_5$ together with a NS three-form $H_3$ and its associated potential $B_2$.

The metric and the RR five form --- the two most relevant quantites for our purposes --- are written  in terms of the vielbein of Eq.~\eqref{Eq-Vielbein} as (with $e^{i j \dots k} \equiv e^{i} \wedge e^{j} \wedge \dots \wedge e^{k}$) 
\be
\label{Eq-RotatedBackgroundMetricF5}
\begin{split}
  ds_{E}^2 & = \sum_{i=1}^{10} (e^i)^2 = \hat{h}^{-\frac{1}{2}} e^{\frac{\Phi}{2}} dx_{1,3}^{2} + k_{1} \hat{h}^{\frac{1}{2}} e^{\frac{\Phi}{2} + 2 k} d\r^{2} + \dots \, , \\
  F_5 & = k_2 \frac{d}{d\r} \left(\frac{e^{2\Phi}}{\hat{h}}\right) \hat{h}^{3/4} e^{-k-\frac{5 \Phi}{4}} \left(-e^{tx1x2x3 \r}+ e^{\theta\varphi 1 2 3} \right) \, .
\end{split}
\ee
The main message up to this point is  the followig. We start with a configuration of D5-branes preserving minimal SUSY. This set-up is characterized by the solutions to what we called the master equation, Eq.~\eqref{Eq-MasterEquation}. Some solutions to this equation can be used to generate other backgrounds like the ones in Eq.~\eqref{Eq-RotatedBackgroundMetricF5} --- with full details given in Appendix~\ref{App-BackgroundDetails}. Regardless of the background being that in Eq.~\eqref{Eq-WrappedD5Metric} or that in Eq.~\eqref{Eq-RotatedBackgroundMetricF5}, there is \emph{one} master-equation controlling them. Once we have solved for $P(\r)$, we can study the dynamics of probe D3-branes embedded in the backgrounds of Eqs.~\eqref{Eq-WrappedD5Metric} or \eqref{Eq-RotatedBackgroundMetricF5} --- passing from one to the other is just a matter of 
appropriately 
choosing the constants $k_1$, $k_2$. Therefore, we will find useful to present a classification of the solutions to Eq.~\eqref{Eq-MasterEquation} in the next section.

\subsection{Solving the master equation: walking solutions}
\label{Sec-WalingSolutions}

We now characterize the properties of the solutions of Eq.~\eqref{Eq-MasterEquation}. Most of the results of this Section were already discussed in \cite{Nunez:2008wi,HoyosBadajoz:2008fw}, hence we will be sketchy.

Since the master equation is a second order differential equation, the general solution to Eq.~\eqref{Eq-MasterEquation} will have two integration constants. In a regime in which $P \gg Q$ (we will discuss its physical meaning below), the master equation is approximately solved by
\be
\label{Eq-SeedSolution}
\tilde{P} = c \left[\cos^3\alpha+\sin^3\alpha \, \left(\sinh (4\r)-4\r\right)\right]^{1/3} \, ,
\ee
which depends on the two integration constants $\alpha$ and $c$. By inspecting the previous expression, one notices that it is possible to determine a scale $\rs$ such that the solution for $P$ is approximately constant for $\r < \rs$, while for $\r > \rs$ behaves as 
$P \propto e^{\frac{4\r}{3}}$. This intermediate scale $\rs$ will play an interesting role below.
In the limit of small $\alpha$, one finds that approximately \cite{Nunez:2008wi}
\be
  4 \rs \simeq \log\left(2\cot^3\alpha\right) \, .
\ee
This allows us to rewrite
\be
\label{Eq-WalkingTildeP}
\tilde{P} \sim c  \sin\alpha  \Big[ e^{4\rs} + 2 (\sinh(4\r)-4\r)\Big]^{1/3} \, .
\ee
Starting from this observation, a more precise characterization of the solution for $P$ proceeds as  follows.

In the far UV, for $\r\rightarrow \infty$, one can check that the following expression solves Eq.~\eqref{Eq-MasterEquation}
\be
\label{Eq-WalkingPUV}
P  =  c_{+} e^{4\r/3} \, + \, 4 \frac{N_{c}^{2}}{c_{+}} \left( \r^{2} - \r + \frac{13}{16} \right) e^{-4 \r/3} \,+ \, \left( -\frac{8}{3} c_{+} \r - \frac{3 c_{-}}{64 c_{+}^{2}} \right) e^{-8\r/3} \, + \, \mcO(e^{-4\r}) \, ,
\ee
where $c_{\pm}$ are the two constants characterizing these solutions. While it will be clearer below, notice that by comparing this with the large-$\r$ limti of Eq.~\eqref{Eq-WalkingTildeP} we can identify $c_{+} \sim c \, \sin\a$.

In the deep IR, for $\r \rightarrow 0$, one finds
 \be
 \label{Eq-WalkingPIR}
 P = c_{0} \, + k_{3} c_{0} \r^{3} + \frac{4}{5} k_{3} c_{0} \r^{5} - k_{3}^{2} c_{0} \r^{6} + \frac{16 (2 c_{0}^{2} k_{3} - 5 k_{3} N_{c}^{2})}{105 c_{0}} \r^{7} \, + \, \mcO(\r^{8}) \, ,
\ee
where now $c_0$ and $k_3$ are the free parameters.

\begin{figure}[ht!]
\begin{center}
\centerline{
\includegraphics[scale=0.8]{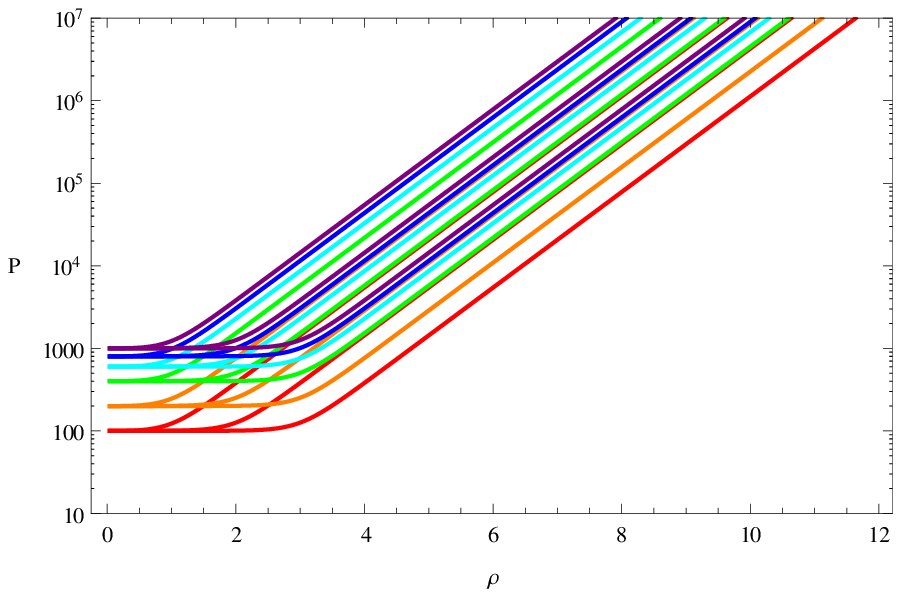}
\hspace*{0.2cm}
\includegraphics[scale=0.8]{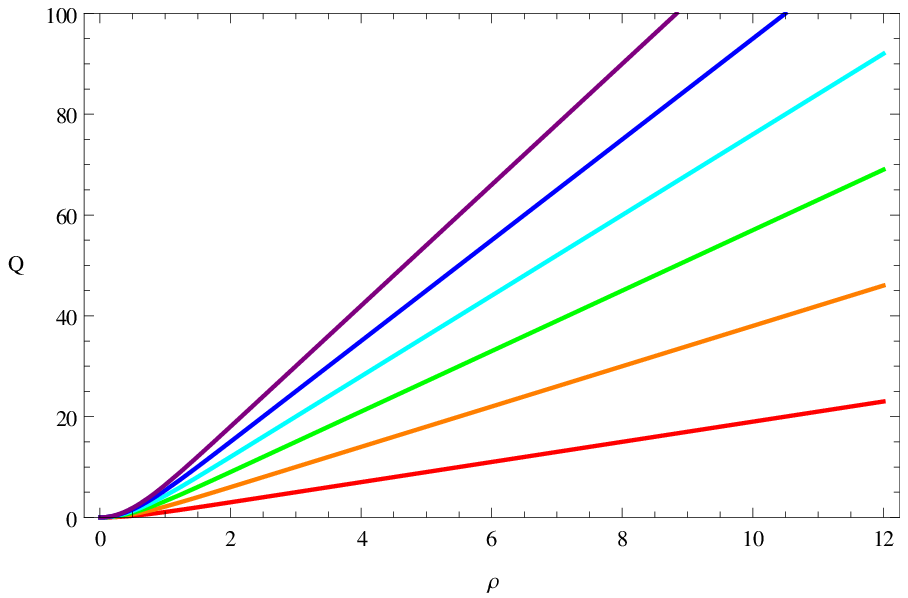}}
\caption{Plots of $P$ and $Q$ as a function of $\r$ for different values of $c_{0}$, $\rs$ and $N_{c}$. The color coding in the left figure is as follows: curves with the same color have a common value of $c_{0}$ between $100$ and $1000$ but differ in their value of $\rs \simeq 1, 2$ or $3$, while $N_{c} = 1$ is always kept fixed. For the right figure, each color corresponds to a different value of $N_{c}$ from $1$ to $6$.}
\label{Fig-Walking-PQ}
\end{center}
\end{figure}

\begin{figure}[ht!]
\begin{center}
\centerline{
\includegraphics[scale=0.6]{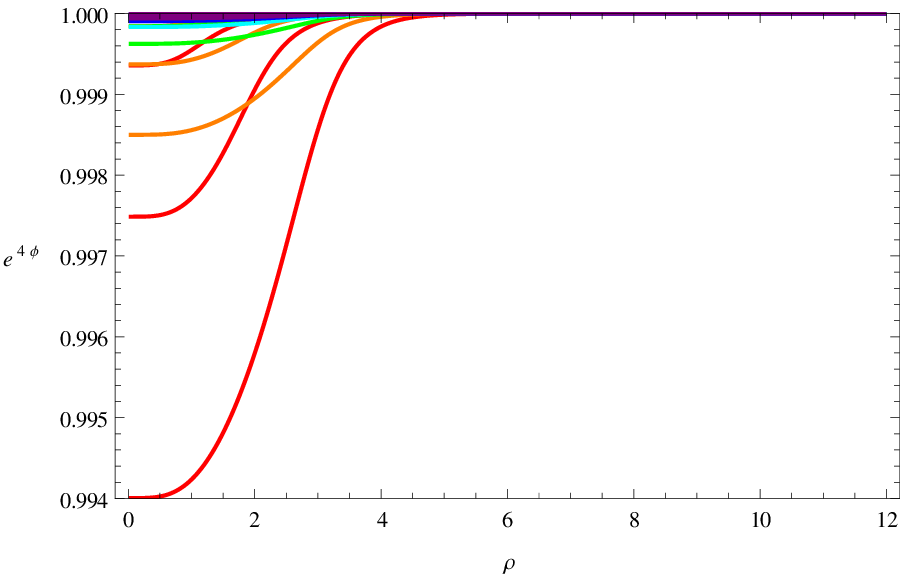}
\hspace*{0.2cm}
\includegraphics[scale=0.6]{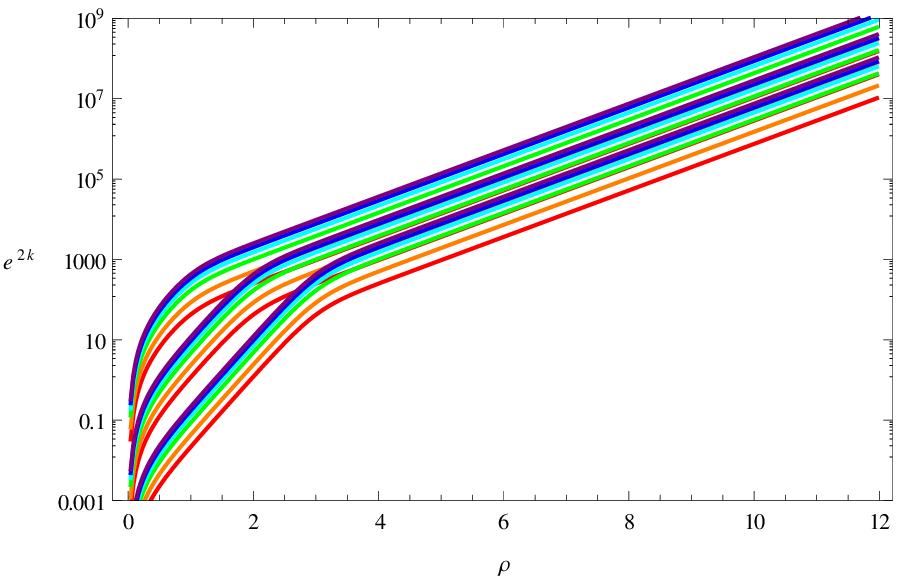}
\hspace*{0.2cm}
\includegraphics[scale=0.6]{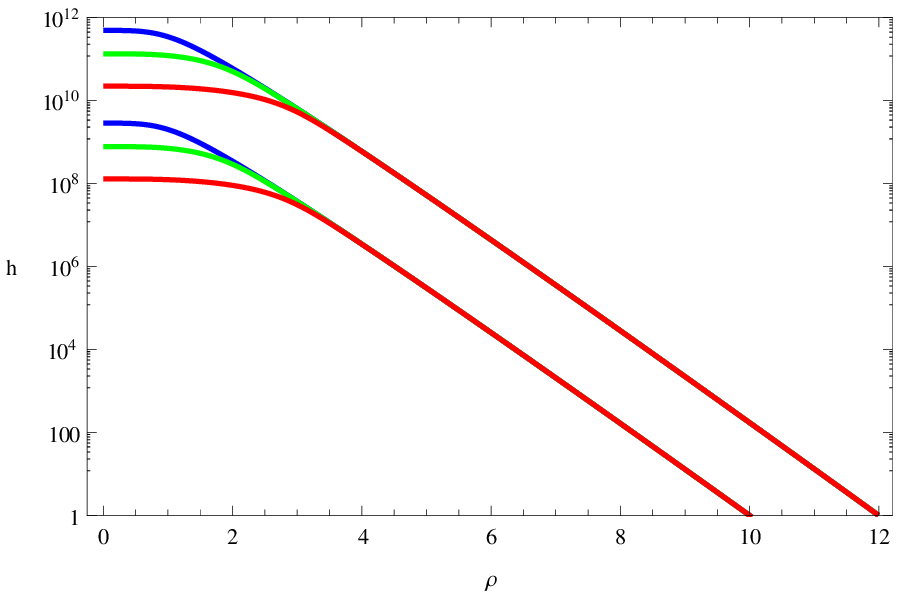}
}
\caption{Plots of $e^{4 \Phi}$, $e^{2 k}$ and $\hat{h}$ as a function of $\r$ for different values of $c_{0}$ and $\rs$ with $N_{c}$ fixed. The color coding is as follows. In the left and middle figure, curves with the same color have a common value of $c_{0}$ between $100$ and $1000$ but differ in their value of $\rs \simeq 1, 2$ or $3$, while $N_{c} = 1$ is always kept fixed. In the right figure, curves with the same color have the a common value of $\rs \simeq $ $1$ (blue), $2$ (green) or $3$ (red) but have a different value of $\ruv = 10$ or $12$ at which we impose the normalization condition $\hat{h}(\ruv) = 1$, see Section~\ref{Sec-PhysCharactGeometry}.}
\label{Fig-Walking-Exp4PhiExp2kWarpFactor}
\end{center}
\end{figure}

One can hence write all of these solutions by 
specifying $N_c$ and for instance any of the pairs $(c, \a)$, $(c_{+}, c_{-})$ or $(c_{0}, k_{3})$, imposing that the function $P$ smoothly connects between IR and UV. Which parameterization to use is mostly a matter of convenience. We found that for analytic computations a particularly convenient choice is $(c_{+}, \rs)$, while for numerically solving the master equation we instead 
specify $(c_{0}, \rs)$. We show some numerical solutions to the master equation for the function $P(\rho)$ 
as well as the function $Q(\rho)$ in Fig.~\ref{Fig-Walking-PQ} and the corresponding background functions $e^{4 \Phi(\r)}, e^{2 k(\r)}$ and $\hat{h}(\r)$ in Fig.~\ref{Fig-Walking-Exp4PhiExp2kWarpFactor}. Note that the warp factor $\hat{h}(\r)$ has significant warping only until $\r \sim \rs$.

To continue further, we need some relation between the UV and IR expansion coefficients. In \cite{Elander:2011mh}, some approximate relations between the coefficients $c_{\pm}, c_{0}, k_{3}$ of the UV and IR expansion and $c$, $\a$ of the approximate solution in Eq.~\eqref{Eq-SeedSolution} were derived. We can use them to express $c_{0}, k_{3}$ and $c_{-}$ in terms of $c_{+}$ and $\rs$ (at leading order in an expansion in large $c_{+}$ and $\rs$):
\beq
\label{Eq-RelUVIRParam}
  c_{0} \sim \displaystyle c_{+} \, e^{4 \rs/3} \quad , \quad c_{-} \sim \displaystyle -\frac{64}{9} \, c_{+}^{3} \, e^{4 \rs} \quad , \quad k_{3} \sim \displaystyle \frac{64}{9} e^{-4 \rs} \left(1 + \frac{8 \, N_{c}^{2} \, e^{-8 \rs/3} \rs^{2}}{ c_{+}^{2}} \right) \, .
\eeq
These relations are  a good approximation if $\frac{N_{c}}{c_{+}} < \displaystyle \frac{3 \, e^{4 \rs/3}}{2^{2/3} \, \rs} \,$.

More systematically, we can describe a solution to the master equation \eqref{Eq-MasterEquation} using a perturbative approach. The solution is then formally given in terms of an infinite series of non-explicit integrals. The usefulness of this formal method will become clear in what  follows. Since this strategy for solving the equation was developed in \cite{Nunez:2008wi} --- see also the Appendix B of \cite{HoyosBadajoz:2008fw} --- we will just summarize here the main points. Note that this approach is essentially a systematic expansion around the (approximate) solution in Eq.~\eqref{Eq-WalkingTildeP}.

We can integrate twice the master equation \eqref{Eq-MasterEquation}, taking already into account the UV and IR asymptotics described above, to get
\be
\label{Eq-MasterEqIntegrated}
  P^{3} - 3 \, Q^{2} P + 6 \!\int_{0}^{\r} \!d\tilde{\r} \, P Q Q' - 12 \!\int_{0}^{\r} \!d\tilde{\r} \, \sinh^{2}(2\r)\, \int_{\tilde{\r}}^{\infty} \! d\hat{\r} \,  \frac{ P' Q Q' }{\sinh^{2}(2\r)} = 16 \, c_{+}^{3} \int_{0}^{\r} \! d\tilde{\r} \, \sinh^{2}(2\r) + P(0)^{3} \, ,
\ee
The double integration provides two integration constants, $P(0)$ and $c_{+}$. The latter is related to the UV behaviour of $P(\r)$ we discussed above, $P(\r) \simeq c_{+} e^{4 \r /3} + \dots \, $. Whilst $P(0)$ is related to the IR behaviour of $P(\r)$. For the walking solutions considered here, we have $P(0) = c_{0} = c_{+} \, e^{4 \rs / 3}$.

\begin{figure}[ht!]
\begin{center}
\centerline{
\includegraphics[scale=0.8]{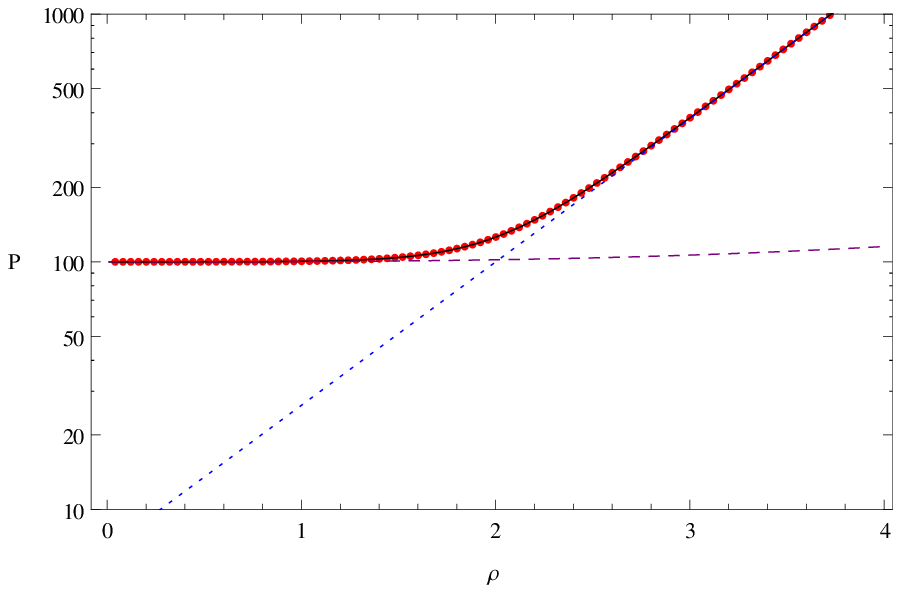}
\hspace*{0.4cm}
\includegraphics[scale=0.8]{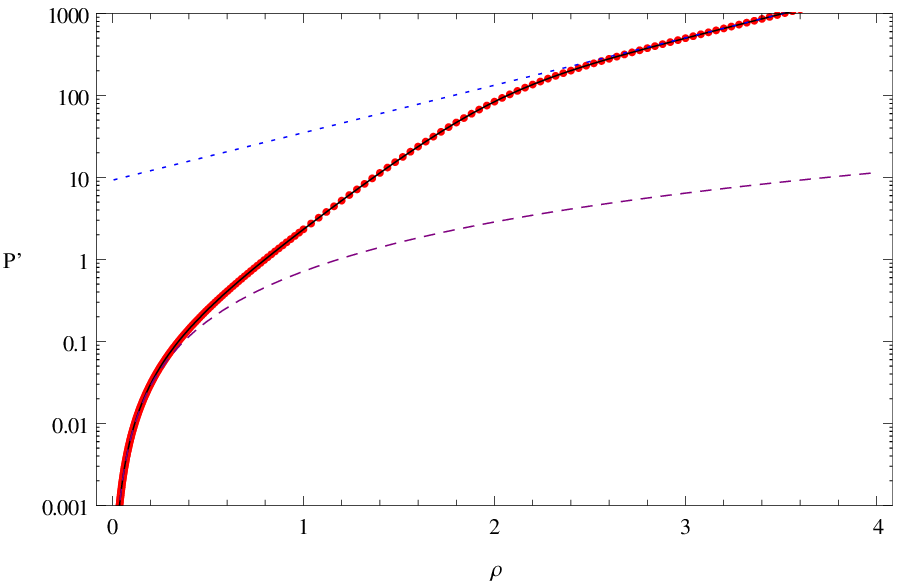}
}
\caption{Plots of $P$ and $P'$ as a function of $\r$ for $N_{c} = 1, c_{0} = 100$ and $\rs \simeq 2$. We show the result of solving the master equation numerically (red dots) as well as three approximations using either only the leading term of the $c_{+}^{-1}$ expansion $P \simeq c_{+} P_{1}$ (black), the leading term of the UV expansion $P \simeq c_{+} e^{4 \r /3}$ (blue dotted) or the leading piece of the IR expansion $P \simeq c_{0} + c_{0} \, k_{3} \, \r^{3}$ (purple dashed).}
\label{Fig-Walking-ApproxP}
\end{center}
\end{figure}

We can formally solve Eq.~\eqref{Eq-MasterEqIntegrated} in a series expansion in $c_{+}^{-1}$. We propose the ansatz
\beq
\label{Eq-SeriesP}
  P(\r)  = \sum_{n=-1}^{\infty} c_{+}^{-n} P_{-n} = c_{+} P_{1} + P_0+ \frac{1}{c_{+}} P_{-1}+\frac{1}{c_+^2}P_{-2} + \dots \, .
\eeq
Inserting this ansatz into Eq.~\eqref{Eq-MasterEqIntegrated} and matching order by order in $c_{+}$ one finds that $P_0=P_{-2}=....=P_{-2k}=0$ and,
\begin{align}
  P_{1} & = \left( 2\left( \sinh(4\r) - 4\r \right) + e^{4 \rs} \right)^{1/3} \, , \label{Eq-SolP1} \\
  P_{-1} & = - \frac{1}{P_{1}^{2}} \left( - P_{1} Q^{2} + 2 \!\int_{0}^{\r} \!d\tilde{\r} \, P_{1} Q Q' - 4 \!\int_{0}^{\r} \!d\tilde{\r} \sinh^2{2\tilde \rho} \int_{\tilde{\r}}^{\infty} \!d\hat{\r} \, \frac{ P_{1}' Q Q'}{\sinh^2{2 \hat \rho}}\right) \, , \label{Eq-SolPm1}
\end{align}
where we have used $P(0)^{3} = c_{0}^{3} = c_{+}^{3} \, e^{4 \rs}$ as indicated in Eq.~\eqref{Eq-RelUVIRParam}. Note again that in Eq.~\eqref{Eq-WalkingTildeP} we can identify $c_{+} \sim c \, \sin\a$. Note also that up to the factor $c \, \sin \a \sim c_{+}$ $P_{1}$ is nothing but $\tilde{P}$ from Eq.~\eqref{Eq-WalkingTildeP}. A recurrence relation for $P_{-(2k+1)}$ can be found by setting $N_{f} = 0$ in Eq.~(B.7) of \cite{HoyosBadajoz:2008fw}. This series converges rapidly to the numerical solutions of Eq.~\eqref{Eq-MasterEquation} --- see Fig.~\ref{Fig-Walking-ApproxP}. However, the information contained in Eqs. \eqref{Eq-SolP1} and \eqref{Eq-SolPm1} is already sufficient for our needs.

\smallskip

Before we close this section, let us briefly elaborate on the physical meaning of the approximate solution in the regime $P \gg Q$, which is given by $\tilde{P}$ in Eq.~\eqref{Eq-SeedSolution}. This form of the solution corresponds to focusing on the limit of large $c_+$, neglecting the terms weighted by inverse powers of this quantity in the expansion Eq.~\eqref{Eq-SeriesP}, \ie dropping all but the first term. It was shown in \cite{Gaillard:2010qg} that in such a limit the geometry of the internal space becomes close to the deformed conifold (or variations of it once we turn on the constant $e^{\rs} \sim c^3 \cos^3\alpha$). This scaling comes together with the dilaton approaching a constant value throughout the entire throat. In other words, for large but finite values of $c_{+}$ we have a background which is `close' to KS 
but not exactly equal to it. As we will see, this leads to a situation in which the force on a probe D3-brane vanishes as $c_{+} \rightarrow \infty$ since the latter does not break SUSY in pure KS (see also Eq.~(15.3) of \cite{Dymarsky:2005xt}).

Let us also briefly comment on the singularity structure of the walking solutions. The backgrounds where $e^{4 \rs} > 0$, do indeed contain a singularity at $\r = 0$. This singularity is `mild' in the sense that the Ricci scalar $R$, the scalar $R_{ab} R^{ab}$ are finite, but the Kretschmann scalar $R_{abcd} R^{abcd}$ diverges. Nevertheless, this singularity does \emph{not} affect our scenario or the results derived from it. As we will discuss below, the inflationary dynamics take place in the interval $[ \r_{I}, \rs ]$, far away from the singularity. Hence, our treatment is free from pathologies and we are calculating in a set-up which is reasonable within the supergravity approximation.

The reader may have observed that in this presentation of the solutions to the master equation \eqref{Eq-MasterEquation} and its rotation to the configuration in Eq.~\eqref{Eq-RotatedBackgroundMetricF5}, various integration constants have appeared: $[k_{1}, k_{2}, \Phi_{0}, c_{+}, \rs]$ are the ones relevant for the topics of this paper. Other constants --- like the choice of origin for the radial coordinate or a constant appearing in the function $Q(\r)$ of Eq.~\eqref{Eq-NewFunctions2} --- either have uninteresting physical meanings, or have been fixed to values that avoid nasty singularities in the geometry. See \cite{Elander:2011mh} for a complete treatment of all integration constants. It is thus useful to characterize the solutions in terms of the relevant constants, imposing some physical criteria to fix some of them. We now turn to discuss this topic.

\subsection{Physical characterization of the geometry}
\label{Sec-PhysCharactGeometry}

The warped throat geometries described above have various features that make them interesting for applications to particle physics model building and, in the present context, for cosmology. Under certain situations, the integration constants characterizing the geometry have a counterpart in a dual QFT, where they can be interpreted as VEVs or couplings of operators perturbing a conformal fixed point. Here, we briefly review the most relevant properties for our purposes, referring the reader to \cite{Nunez:2008wi,Elander:2011mh} for more details.

If the integration constant $\rs$ is sufficiently large and positive, there is an intermediate regime in the radial direction, from $\r_{I} \sim 1$ to $\r \sim \rs$, in which the effective 4d gauge coupling of the dual field theory is finite and approximately constant \cite{Nunez:2008wi}. Hence, the geometry associated with solutions with large $\rs$ are suitable for describing field theories exhibiting \emph{walking} dynamics.  In  the QFT dual, the VEVs of two operators start to dominate in the regime
$\r_{I} \le \rho \le \rs$, which are absent in the KS configuration: a dimension-2 VEV that brings the background on the baryonic branch \cite{Butti:2004pk}, and a dimension-6 VEV associated with the walking dynamics. The combined action of the two operators, enriches the dynamics generating the intermediate walking region  and provides yet another physical scale, at the value  $\rho_I \sim 1$ in the radial coordinate $\rho$, corresponding  approximately to the scale where the walking regime ends and the system enters into a confining regime. At the geometrical level,
the position $\r_{I}$ roughly corresponds to the value of the radial coordinate below which the functions $a(\r)$, $b(\r)$ (controlling $F_{3}, H_{3}$) defined in Appendix~\ref{App-BackgroundDetails} become non-trivial --- this fact is associated with the spontaneous breaking of a discrete $\mathbb{Z}_{2 N_{c}}$ global symmetry and with a confining behaviour --- see the second paper of \cite{Elander:2009pk}. As we will see in what follows, the region $\r_{I} \lesssim \rho \lesssim \rs$ is important for building our inflationary set-up.

 \smallskip
 
In order to consider inflationary models, we need to glue our throat geometry to a compact space, say a (warped) CY-three-fold, in the UV. To this extent, let us discuss the UV behavior of $\hat{h}(\r)$, which is controlled by the constant $k_{2}$. For generic values of the parameter $k_2$, the warp factor $\hat{h}(\r)$ asymptotes to a constant as the radial coordinate $\r$ grows. In the field theory dual, this is associated with the presence of a dimension-8 operator that dominates the dynamics in the UV and needs a careful specification of a UV completion --- see the discussion in \cite{Elander:2011mh}. In this UV region, the geometry generically differs considerably from a space-time dual to a conformal theory. Thus, it makes little sense to interpret the duals of the integration constants of the geometry as operators representing `small perturbations' around a conformal fixed point. Fortunately, the UV behavior can be improved: tuning the parameter $k_2$ appearing in the warp factor $\hat{h}$, cf. Eq.~\eqref{Eq-RotatedWarpFactor}, to be $k_{2} = e^{-\Phi(\infty)}$ adiabatically switches off the aforementioned dimension-8 operator, and the geometry becomes (logarithmically) close to AdS for large values of the radial coordinate \cite{Elander:2011mh}, thanks to  RR five-form and NS three-form fluxes that get turned on 
at $\rho\simeq \rs$  ---  see the plots in \cite{Elander:2011mh}. 
The configuration at $\r \to \infty$ becomes \emph{almost} identical to the Klebanov-Strassler background \cite{Klebanov:2000hb}. 
So by making this choice for $k_2$, we can reliably identify the operators in 
the dual field theory representing `small perturbations' from a quasi-conformal regime. 
This is good news also for the aim to build an inflationary model. After tuning $k_2$ as explained above, in the UV, 
the throat is very similar to AdS space, which implies that upon gluing the throat into a compact space the 
(four-dimensional) graviton zero mode will be mostly localized in the compact part of the bulk and not in the warped throat. This information will be useful in what follows, to fix the value of the effective four-dimensional Planck mass in string units. Henceforth, we always set $k_2\,=\,e^{-\Phi(\infty)}$.

\smallskip 

Let us proceed with specifying the UV properties of our configuration. In the limit $c_+ \to \infty$, the dilaton becomes close to a constant, see for example \cite{Gaillard:2010qg,Elander:2011mh}. Without lack of generality, we can choose boundary conditions such that in the far UV $\Phi(\infty) \equiv 0$, such that $k_2=1$. Consider the expression for the dilaton in Eq.~\eqref{Eq-NewFunctions2}; the asymptotic expansion for $e^{4\Phi}$ as $\r \rightarrow \infty$ reads
\be
\label{Eq-DilatonUV}
  e^{4 \Phi - 4 \Phi_{0}} \simeq \frac{3}{8 c_{+}^{3}} \left( 1 - \frac{3 \, N_{c}^{2} \, e^{-8 \r/3} \, (8 \r - 1)}{4 \, c_{+}^{2}} + \dots \right) \, .
\ee
Given the requirement above, $\Phi(\infty) = 0$, the parameter $\Phi_{0}$ is then determined in terms of $c_{+}$ as
\beq
\label{Eq-FixedPhi0}
  \Phi_{0} = \frac{1}{4} \ln \left( \frac{8}{3} \, c_{+}^{3} \right) \,\, \Leftrightarrow \,\, e^{4 \Phi_{0}} = \frac{8}{3} \, c_{+}^{3} \, .
\eeq

Another physically convenient requirement allows to fix the parameter $k_1$. The warp factor $\hat{h}$ is (we set $k_2=1$),
\beq
  \hat{h} = k_{1}^{-2} (1 - e^{2 \Phi}) \, .
\eeq
We choose the value $k_1$ such that at the UV boundary on which the throat is joined to the compact space the warp factor is of order one, $\hat{h}(\ruv) \sim 1$. This implies that the size of the compact bulk is determined by the string scale, and its volume is not enhanced by large values of the overall warp factor. This was also done in \cite{Kachru:2003sx} and is important for fixing the scales that enter in the phenomenological discussion of the inflationary set-up we are going to develop.

Using Eqs.~\eqref{Eq-DilatonUV} and \eqref{Eq-FixedPhi0}, the condition $\hat{h}(\ruv) = 1$ implies
\beq
\label{Eq-Fixedk1}
  k_{1} \simeq \sqrt{\frac{3}{8}} \, \frac{N_{c} \, e^{- 4 \r_{\UV} / 3} \, \sqrt{8 \r_{\UV} - 1}}{ c_{+}}\,.
\eeq
After making this choice, we have fixed three of the parameters controlling the warped throat $k_1, k_2, \Phi_0$. We still can vary $c_+$ (controlling the profile of the dilaton field), $\rs$, the position $\ruv$ at which the throat ends and the number $N_c$ of wrapped D5-branes at the tip of the throat.

\smallskip

\begin{figure}[ht!]
\begin{center}
\centerline{
\includegraphics[scale=0.6]{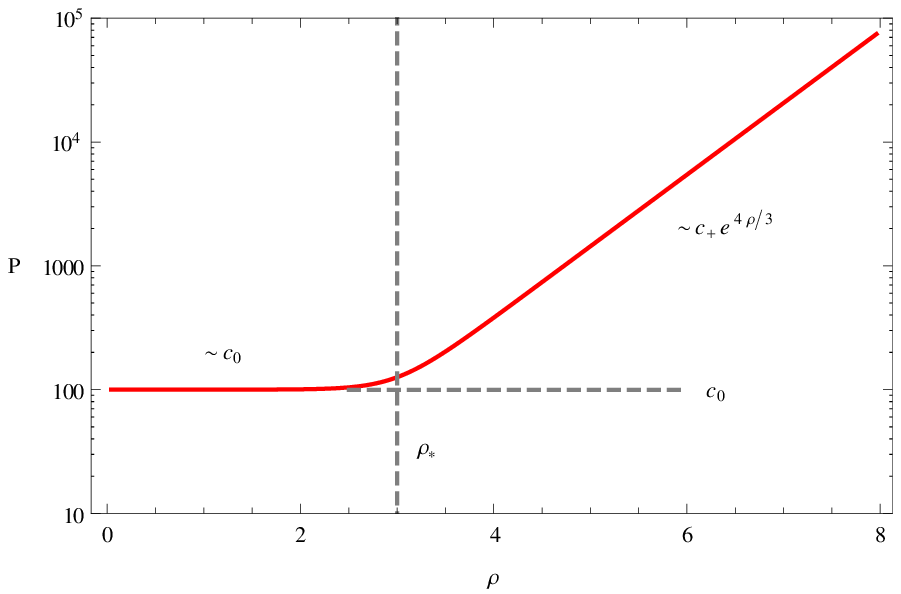}
\hspace*{0.2cm}
\includegraphics[scale=0.6]{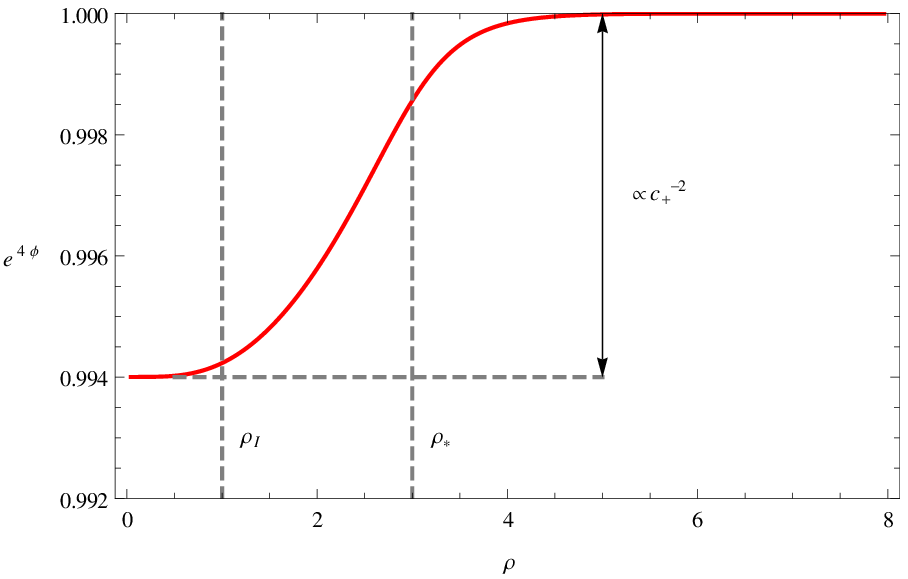}
\hspace*{0.2cm}
\includegraphics[scale=0.6]{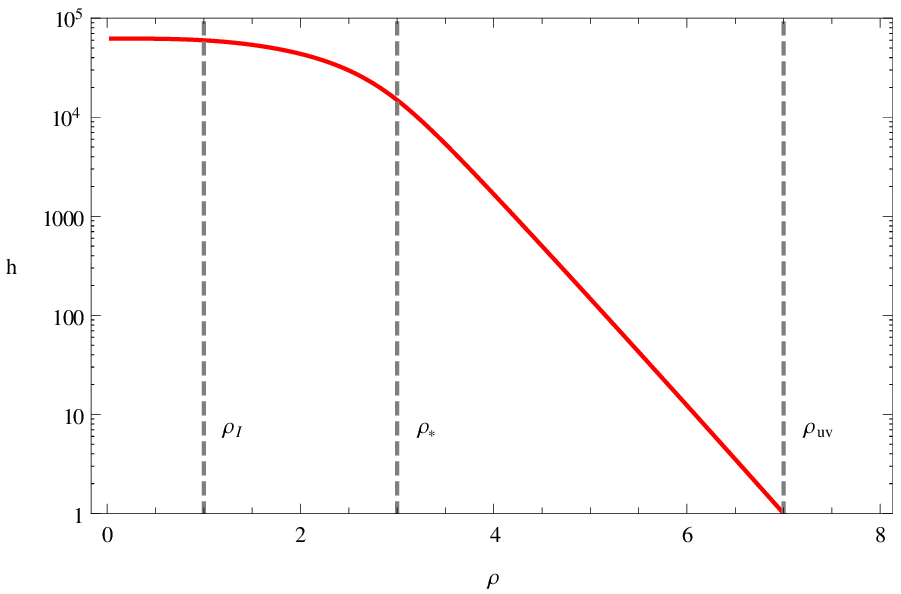}
}
\caption{Plots of $P, e^{4 \Phi}$ and $\hat{h}$ as a function of $\r$ for $N_{c} = 1$, $c_{0} = 100$, $\rs \simeq 3$ and $\ruv = 7$. The different scales $\rs$, $\r_{I}$ and $\ruv$ playing a role in this paper are indicated by the gray dashed lines.}
\label{Fig-Walking-MarkedPExp4Phi}
\end{center}
\end{figure}

All these parameters and scales discussed above will characterize our model of inflation. 
To close this section, let us briefly summarize the most important parameters specifying the geometry and their physical interpretation, see also Fig.~\ref{Fig-Walking-MarkedPExp4Phi}.
\begin{itemize}
  \item{ {\bf $c_{+}$:} controls VEV of a dimension-2 operator $\langle \mcU_{2} \rangle \sim \frac{N_{c}}{c_{+}}$ corresponding to motion along baryonic branch --- it also
induces dilaton profile.}
  \item {{\bf $\rs$:} controls VEV of a dimension-6 operator $\langle \mcO_{6} \rangle \sim e^{4 \rs}$ inducing walking dynamics --- also modifies the dilaton profile, especially in IR region.}
  \item {\bf $\ruv$:} controls size of warp factor at the tip, \ie ratio between confining scale and UV cutoff ---  distance over which strong warping occurs: $\sim \ruv - \rs \,$.
\end{itemize}

\section{Brief review of warped D3 inflation}
\label{Sec-ReviewD3Inflation}

We consider a probe D3-brane placed in a warped throat geometry. The inflaton field will geometrically correspond to the radial position of the D3-brane on the throat (the first papers identifying the inflaton with brane motion have been \cite{Dvali:1998pa,Dvali:2001fw,Burgess:2001fx}). To derive the canonically normalized inflationary action, 
we start from a generic string frame metric --- one can show that 
the relevant formulae 
below have the same expression if we work in Einstein frame --- 
with a warped conical region and RR four-form ansatz of the form
\beq
\label{Eq-MetricC4Ansatz}
ds_{st}^2 = H_1 \, dx_{1,3}^2 +  H_2 \, d\r^2 + \dots \, , \qquad C_4= \, \mcC_4(\r) \, dt\wedge dx_1 \wedge dx_2 \wedge dx_3 \, ,
\eeq
where the dots denote the angular part of the metric. The induced metric on a D3-brane that extends in the four non-compact directions and moves along the radial direction,
\beq
\Sigma_4=[t,x_1, x_2, x_3] \, , \quad \r = \r(t) \, ,
\eeq
is given by
\beq
ds_{ind}^2= H_1 \, (dx_1^2 + dx_2^2 + dx_3^2) + (H_2 \, \dot{\r}^{2} - H_1) \, dt^2 \, .
\eeq
The BIWZ action for this probe brane is, again in string frame, 
\beq
\begin{split}
S_{BIWZ} & = -T_{3}\int d^4 x\Big( e^{-\Phi}\sqrt{-\det[g_{ind}]}- \, \mathcal{C}_4 \Big) \\
  & = -T_{3}\int d^4 x\Big(e^{-\Phi}H_1^2 \sqrt{1-\frac{H_2}{H_1}\dot{\r}^2}-  \, \mathcal{C}_4 \Big) \, .
  \label{Eq-ProbeBraneAction}
\end{split}
\eeq
We expand for small velocities $\dot{\r}$ and find
\beq
S_{BIWZ}= T_{3} \int d^4 x \left( \frac{H_2 H_1 e^{-\Phi}}{2}\dot{\r}^2 - e^{-\Phi}H_1^2 \left(1- \frac{\k \, e^{\Phi} \, \mathcal{C}_4}{H_1^2} \right) \right) \, ,
\eeq
from which we read off the effective D3-brane potential
\beq
V = T_{3} \, e^{-\Phi} \, H_1^2 \, \left(1- \frac{ \, e^{\Phi} \, \mathcal{C}_4}{H_1^2} \right) \, .
\eeq
The canonical radial variable, in string frame, is determined from 
\beq
\label{Eq-CanonicalRadialCoord1}
dr = \sqrt{T_3}\,e^{-\Phi/2} \, \sqrt{H_1H_2} \, d\r \, .
\eeq
The determination of the potential for the D3-brane may then proceed following  a two-step procedure. In a first step, we can find supersymmetric solutions for the background field equations, as we did in the previous section, and embed a probe D3-brane in the corresponding configuration. If the brane breaks SUSY, it feels a force that makes it move. This motion induces a modification of the warped geometry, which we assume to be limited to change the 4d metric $dx_{1,3}^{2}$ describing the observed four dimensions, lying inside the warp factor in Eq.~\eqref{Eq-WrappedD5Metric} --- or in its `rotated' version in Eq.~\eqref{Eq-RotatedBackgroundMetricF5}. Namely, we assume that this 4d metric changes from Minkowski to an homogeneous and isotropic FRW space time. For this approach to be valid, we make the hypothesis that all the relevant moduli are stabilized with sufficiently high masses such that they are not destabilized by the brane motion. 
The second step \cite{Baumann:2008kq,Baumann:2009qx,Baumann:2010sx} consists of starting with a fiducial warped throat configuration, for example the KS throat, on which a probe D3 feels no (or only a small) force, and calculating perturbatively the D3-brane potential around this configuration generated by the backreaction on the geometry from corrections sourced \eg by SUSY breaking contributions in the bulk necessary to stabilize the geometric moduli.

In any case, the D3-brane action then has to be supplemented by coupling it to gravity. This is done by gluing the throat to a compact space, which induces an Einstein-Hilbert term completing the D3-brane action. Usually, this gluing also provides new contributions to the D3 brane potential, whose combination is neglected in the first step described above, while it can be computed perturbatively in the second step at least in principle.

The main goal of the first step, which we will pursue in this paper, is to clearly identify the forces that induce the brane motion, offering neat physical and geometrical insights for understanding inflation in geometrical terms, and possibly leading to a fully calculable inflationary potential. The results can also be interpreted in a dual QFT, offering interesting new perspectives on possible mechanisms that drive inflation from the point of view of strongly-coupled field theories. The problem with this first step, is of course that one does not take into account \emph{all} possible contributions to the inflationary potential since contributions depending on couplings with other light moduli or fields in the compact bulk are assumed to be negligible. This is in general not consistent and is generally associated with  the so called $\eta$ problem (more on this in Section~\ref{Sec-Discussion}).

The purpose of the second step is then to  offer  a systematic way to calculate perturbatively the brane potential, offering fully reliable results, although the calculations are not always straightforward, and a further numerical analysis of the inflationary trajectory is often required (see \eg \cite{Agarwal:2011wm,Dias:2012nf,McAllister:2012am} for the case of the KS throat). Ideally, the physical and geometrical insights about the forces acting on the brane within a given set-up, acquired by the first step, should be supplemented by a careful determination of the corrections to the inflationary potential, calculated following the second step. In this way, one can obtain a fully-controlled satisfactory scenario for warped brane inflation. 

\smallskip

After clarifying  these methodological issues, let us specify two quantities --- the slow-roll parameters  $\e$ and $\eta$ --- which are important for characterizing our inflationary model. In the limit in which the brane moves slowly, the slow-roll parameters are given by
\be
\label{Eq-SlowRollParametersRho}
\begin{split}
 \epsilon & = \frac{\mp^{2}}{2 \,V^2} \left(\frac{\partial V}{\partial \r} \right)^2 \,\left( \frac{d \rho }{d r} \right)^2 \, , \\ 
 \eta & = \frac{\mp^{2}}{V } \left[ \frac{\partial^2 V}{\partial \r^2}  \,\left( \frac{d \rho }{d r} \right)^2 
+ \left( \frac{\partial V}{\partial \r} \right)\,\left( \frac{d^2 \rho }{d r^2} \right) \right] \, .
\end{split}
\ee
Note that these are simply the usual slow-roll parameters in terms of the 
canonically normalized radial coordinate $r$ obtained from Eq.~\eqref{Eq-CanonicalRadialCoord1}, namely
\beq
\label{Eq-SlowRollParameters}
 \epsilon = \frac{\mp^{2}}{2V^2} \left( \frac{\partial V}{\partial r}\right)^2 \, , \quad \eta = \frac{\mp^{2}}{V} \frac{\partial^2 V}{\partial r^2}.
\eeq
Given a particular warped geometry and its field content, the resulting inflationary potential built along these lines can exhibit regions that are flat enough for supporting a sufficiently long period of slow-roll inflation. In this set-up, we can interpret the properties of the inflationary scenario using the geometrical features of the throat, or the properties of its dual QFT.


\section{Inflationary case study --- Walking solution}
\label{Sec-SlowWalkingInflation}

Let us now discuss how inflation occurs in the geometry of the walking solution discussed below Eq.~\eqref{Eq-SeedSolution}. In the string frame, the functions $H_{1,2}$, $\mcC_4$ in Eq.~\eqref{Eq-MetricC4Ansatz} for this solution are given by
\beq
  H_{1} = \displaystyle \hat{h}^{-1/2} e^{\Phi} \quad , \quad H_{2} =  k_1 \hat{h}^{1/2} e^{2 k + \Phi} \quad , \quad \mcC_{4} =   \hat{h}^{-1} e^{2\,\Phi}\,k_2 \, .
\eeq

The inflationary potential $V$ and canonically normalized radial variable $r$ (which acts as the inflaton field) are
\beq
\label{Eq-BranePotRadialCoord}
  V = \frac{k_{1}^{2} \, T_{3}}{e^{-\Phi} + k_{2}} \quad \text{and} \quad  dr = \sqrt{T_3\,k_{1}} e^{k + \Phi/2} d \r \, .
\eeq
In the following, we take $k_2=1$ and $k_1$ given by formula \eqref{Eq-Fixedk1} as discussed in Section~\ref{Sec-PhysCharactGeometry}. 

\begin{figure}[ht!]
\begin{center}
\centerline{
\includegraphics[scale=0.6]{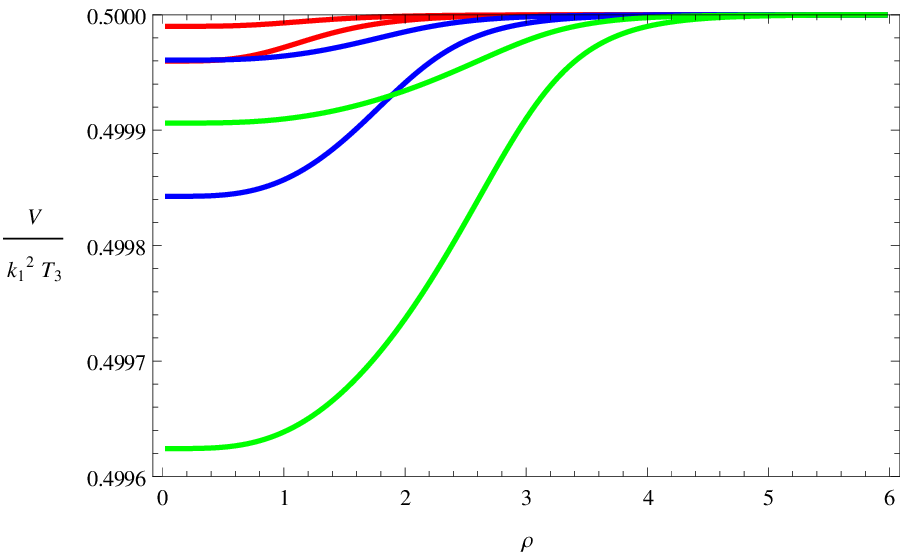}
\hspace*{0.4cm}
\includegraphics[scale=0.6]{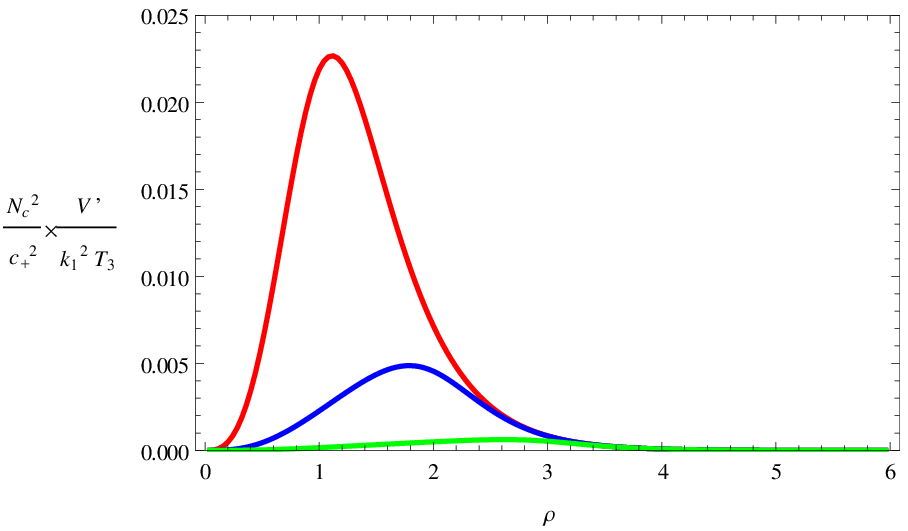}
\hspace*{0.4cm}
\includegraphics[scale=0.6]{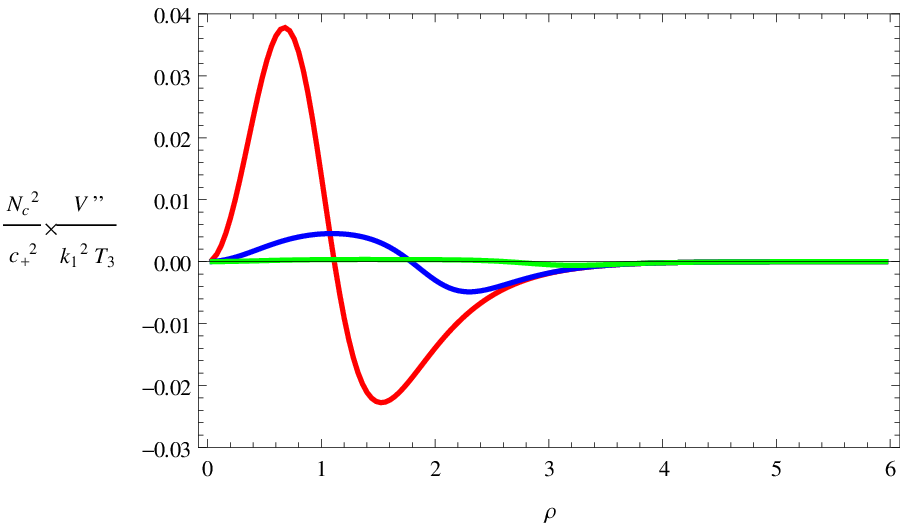}
}
\caption{Plots of $V$, $V'$ and $V''$ as functions of $\r$ rescaled by appropriate factors. The color coding is as follows: each color corresponds to a different value of $\rs \simeq$ $1$ (red), $2$ (blue) or $3$ (green); in the left figure curves with the same color differ by their value of $c_{0} = 100$ or $200$. We always keep $N_{c} = 1$ fixed.}
\label{Fig-Walking-RescaledPotential}
\end{center}
\end{figure}

In the limit $\nicefrac{N_c}{c_+} \to 0$, our geometry becomes (almost) identical to the Klebanov-Strassler throat: it would coincide with KS in this limit, if additionally we consider the limit $\rs \to - \infty$; see Eq.~\eqref{Eq-SolP1}. For $\nicefrac{N_c}{c_+} \to 0$, the dilaton $\Phi$ becomes a constant, and thus the brane does not feel any force. Outside this limit, a non-trivial profile for $\Phi$ forces the brane to move, leading to cosmological evolution compatible with inflation. The main advantage of this framework is that a controllable brane motion is associated with the properties of our background, which is free of singularities in the region of interest, and we do not need to add anti-D3 objects to make the brane move. In this sense, our scenario can be considered as a generalization of the inflationary D-brane model in the baryonic branch of the KS throat \cite{Dymarsky:2005xt}. Let us briefly explain the differences between the two setups. First, in \cite{Dymarsky:2005xt} it was proposed to have inflation in the AdS-like UV region, while in our case inflation takes place near the tip in the IR. Second, as we will see below, in our case $\eta$ becomes small in the vicinity of an inflection point and additionally the walking dynamics lead to a much stronger suppression of $\e$ by a factor of $e^{- 8 \rs /3}$.
 
\smallskip

Before studying in detail the quantitative features of an inflationary model that can be build using this geometry, let us start with some general qualitative considerations. In the deep IR, $\r \to 0$, the geometry is essentially controlled by the wrapped D5-branes. The corresponding warping creates the attractive force felt by the moving D3-brane. Moreover, the fields sourced by the D5s, which are concentrated in the IR region, make this attractive force stronger and stronger when increasing  the radial coordinates. On the contrary, at larger values of $\r$, we are leaving the region in which the wrapped D5-branes dominate, and the warping changes. Additionally, for $\r > \rs$, new fluxes (the RR five-form and NS three-form) appear and change the throat geometry, making it similar to the KS throat. Consequently, the attractive force acting on the brane becomes weaker, until it asymptotically vanishes in the far UV when the geometry quickly approaches KS. In this region, however, the throat will be attached to a compact space as we explained above, and drastic modifications of the background have to be expected in the gluing region. See Fig.~\ref{Fig-Walking-RescaledPotential} for plots of $V$, $V'$ and $V''$ for different values of $c_{0}$ and $\rs$.

On general grounds, we expect a region of transition, in which the attractive force acting on the brane reaches a maximum in size at a certain value of the radial coordinate after which its size starts to decrease (see the middle figure of Fig.~\ref{Fig-Walking-RescaledPotential}). In the following, our aim is to identify this region, and we will focus on it for building a model of slow-roll inflation with the necessary properties to match observations. As we will discuss
in detail, this region is characterized by a potential that is flat enough to provide $60$ $e$-folds of inflation. Remarkably, we will show that our set-up allows to determine and analyze the inflationary potential analytically, at least in a physically well-motivated limit. In the QFT dual, this region is precisely the walking region in which the coupling constant varies only mildly. (Recently, similar considerations have been made in a conceptually different set-up in \cite{Evans:2012jx}.)

\subsection{The D3 brane potential  in the large-$c_+$ and  large-$\rs$ limits}
\label{Sec-ScalingBehaviour}

As explained above, in the large-$c_+$ limit, our geometry approximates the KS throat, and we expect that in this limit the force acting on the D3 becomes small and capable of supporting a sufficiently long period of slow-roll inflation. The limit of large $c_+$ is thus physically interesting. It is moreover technically convenient since the formulae simplify considerably and a fully analytical treatment can be carried out.

\smallskip

Using the series expansion for $P$, cf. Eqs.~\eqref{Eq-MasterEqIntegrated} and \eqref{Eq-SeriesP}, we can write the leading contributions to $P$ as
\be
\label{Eq-SeriesP2}
P =  c_{+} P_{1} + \frac{1}{c_{+}} P_{-1} + \dots \, ,
\ee
where $P_1$, $P_{-1}$ do not depend explicitly on $c_{+}$ and are given in Eqs. \eqref{Eq-SolP1} and \eqref{Eq-SolPm1}. In the following, we ignore all terms suppressed by higher powers of $c_+$.

Inserting Eq.~\eqref{Eq-SeriesP2} into the expression for the dilaton field, we find at leading order in an inverse $c_{+}$ expansion
\be
e^{4 \Phi} = \frac{2 \, e^{4 \Phi_{0}} \,\sinh^{2}(2 \r)}{c_{+}^{3} \, P_{1}^{2} P_{1}'}\,\left[ 1 + \frac{1}{c_{+}^2} \left( \frac{Q^{2}}{P_{1}^{2}} - 2 \frac{P_{-1}}{P_{1}} - \frac{P_{-1}'}{P_{1}'} \right) + \dots \right] \, ,
\ee
where the dots indicate terms suppressed by higher powers of $c_{+}$. Defining
\be
\label{Eq-DefinitionM}
-N_{c}^{2} \, \mcM(\rho) \equiv \frac{Q^{2}}{P_{1}^{2}} - 2 \frac{P_{-1}}{P_{1}} - \frac{P_{-1}'}{P_{1}'}
\ee
and using the identity (which can be easily derived using Eq.~\eqref{Eq-SolP1})
\be
  \frac{2 \,\sinh^2(2 \r)}{P_{1}^{2} P_{1}'} \equiv \frac{3}{8} \, ,
\ee
we can write the D3-brane potential of Eq.~\eqref{Eq-BranePotRadialCoord} as follows:
\be
\label{Eq-BranePotRadialCoord2}
  V =  \frac{3\, N_c^2 \,T_{3}\, (8\ruv-1)\, e^{-8 \ruv/3}}{{16} \, c_{+}^2}\left( 1- \frac{N_{c}^{2}}{8 \, c_{+}^{2}} \mcM(\r) + \dots \right) \, .
\ee

This potential clearly vanishes in the limit of $c_+ \to \infty$. Indeed, in this limit the D3 brane becomes a BPS state (thus experiences no force). The function $\mcM$ in Eq.~\eqref{Eq-DefinitionM}, characterizes the field dependence of the potential. Its features are controlled by the functions $P_{-1}$ and $Q$, quantities that control how much the geometry deviates from KS. Hence, it is clear that the features of the potential  directly depend on the characteristics of the geometry. Using the following identity
\be
\frac{P'_{-1}}{P'_1} = -\frac{2\,P_{-1}}{P_1}+\frac{Q^2}{P_1^2}+\frac{4\,\sinh^2{2  \rho} }{ P_1' \,P_1^2}\,\int_{{\r}}^{\infty} \!d\hat{\r} \, \frac{   P_{1}' Q Q'}{\sinh^2{2 \hat \rho}} \, ,
\ee
which can be derived using Eq.~\eqref{Eq-SolPm1}, we easily obtain an integral expression for $\mcM(\rho)$:
\be
N_c^2\, \mcM(\rho) =  \frac{4\,\sinh^2{2 \rho} }{ P_1' \,P_1^2}\,\int_{{\r}}^{\infty} \!d\hat{\r} \, \frac{ P_{1}' Q Q'}{\sinh^2{2 \hat \rho}} = \frac34 \,\int_{{\r}}^{\infty} \!d\hat{\r} \, \frac{ P_{1}' Q Q'}{\sinh^2{2 \hat \rho}}  = 2\,\int_{{\r}}^{\infty} \!\frac{d\hat{\r}}{P_{1}^2} \,  \frac{ \partial Q^2}{\partial \hat \rho} \, .
\ee
This will be of great use below. 
In particular, the derivative of $\mcM$ reads
\be
N_c^2 \, \mcM'(\rho) = -\frac{2}{P_{1}^2} \, \frac{\partial Q^2}{\partial \rho} \, ,
\ee
implying that the force acting on the brane is always attractive since the derivative of $Q^2$ as obtained from Eq.~\eqref{Eq-NewFunctions2} is always positive.

At leading order in a $c_{+}^{-1}$-expansion, the relation between the canonically normalized radial coordinate $r$ as given by Eqs.~\eqref{Eq-CanonicalRadialCoord1}, \eqref{Eq-BranePotRadialCoord} and the usual coordinate $\rho$ does not depend on $c_{+}$, and reads
\be
\begin{split}
d r & \simeq \sqrt{\frac{N_c T_{3}}{4}} \,\left[\frac{3\left( 8 \ruv-1\right) }{2} \right]^\frac14\, \sqrt{  P_{1}'} \,e^{-\frac23 \ruv}\, d\r \\
 & =\sqrt{2 N_c T_{3}} \,\left[\frac{2\left( 8 \ruv-1\right) }{3} \right]^\frac14\,\frac{\sinh{2 \rho}}{P_1}\,e^{-\frac23 \ruv}\, d\r  \, .
\end{split}
\ee

\bigskip

The formulae become even simpler when focussing on the region $\rho \ll \rs$. This is the region on which we would like to concentrate our attention: precisely in this region the attractive force acting on the brane has a maximum at $\rho \sim \rho_0$, and then starts to
slowly  decrease in size towards larger values of $\r$.

In order to study this feature, we focus on a limit in which $\rs$ (or more precisely $e^{\rs}$) is much larger than one. In this limit, we can write the leading terms in $P_1$ and its derivative as
\begin{align}
  P_{1} & \simeq e^{4 \rs /3} \left[ 1 + \frac{2}{3} e^{-4 \rs} \left( \sinh(4\r) - 4\r \right) \right]\, , \\
  P_{1}' & \simeq \frac{16}{3} e^{-8 \rs/3} \sinh^{2}(2\r) \, ,
\end{align}
Thus, we have
\be
N_c^2 \, \mcM'(\rho)\, \simeq \,-2\,e^{-\frac{8}{3} \rs} \, \frac{\partial Q^2}{\partial \rho} \, ,
\ee
and
\be
d r  \simeq 2 \sqrt{\frac{N_c T_{3}}{3}} \,\left[\frac{3\left( 8 \ruv-1\right) }{2} \right]^\frac14\,  \,e^{-2 \ruv/3}\, e^{-4 \rs /3} \, \sinh (2\rho) \, d\r \, .
\ee
At this point, we have all the tools necessary to analyze the maximum for the attractive force acting on the brane. This corresponds to a point in which the second derivative of the brane potential $V$ along the canonically normalized radial coordinate $r$ vanishes: it is an \emph{inflection point} of the brane potential.

A straightforward calculation shows that such a point exists, and is (approximately) determined by the root of the following equation in the coordinate $\rho$:
\be
\label{Eq-InflectionPointLocation}
  -5 + 40 \rho^2 + 4 (1 + 6 \rho^2) \cosh{4 \rho} + \cosh{8 \rho} - \rho \, (30 \sinh{4 \rho} + \sinh{8 \rho}) = 0 \, .
\ee
Remarkably, this equation does not involve any of the parameters defining the geometry and is valid in the limit in which both $c_+$ and $e^{\rs}$ are much larger than one. Numerically, one finds that the root of the previous equation is at $\r_{0} \simeq 0.98$ --- we have checked using numerical solutions that large but finite values of $c_{+}$ and $\rs$ only lead to tiny shifts in the position of $\r_{0}$. For $\rho \ge \rho_0$, the attractive force acting on the brane starts to decrease in size. In the following, we will focus on the region $\rho_0 \le \rho\ll \rs$ in our search for an inflationary trajectory. The effective potential depends on the tunable parameters $c_+$, $\rs$, $\ruv$ that allow us to change both its size and shape. Geometrically, they control the profiles for the dilaton, warp factor and the $p$-form fields. These parameters will provide considerable flexibility for tuning the properties of our inflationary model. A qualitative understanding of the shape of the force acting on the probe D3-brane using aspects of the dual QFT 
is postponed to Appendix~\ref{App-CommentsDualQFT}.

\subsection{Inflationary quantities and their scaling behavior}

In the limit of large $c_+$ and $\rs$, we can obtain \emph{analytic} expressions in the parameters controlling the inflationary dynamics. These analytic expressions are simple enough to enjoy a clean physical interpretation.

\smallskip

\noindent

$\bullet$ {\bf Slow-roll parameters:} Using the expression for the D3 brane potential in Eq.~\eqref{Eq-BranePotRadialCoord2}, it is straightforward to calculate the slow-roll parameters. At leading order, the large-$c_+$ expansion of the $\epsilon$-parameter in Eq.~\eqref{Eq-SlowRollParametersRho} reads
\beq
  \e \simeq \frac{\sqrt{3}}{{64} \sqrt{2}} \, \frac{\mp^{2}}{T_3} \, \frac{ e^{4 \ruv/3}}{ N_c c_{+}^{4} \sqrt{8 \ruv - 1}} \, \left( \frac{1}{P_{1} \sinh(2\r)} \, \frac{\6 Q^2}{\6 \r} \right)^2 \, .
\eeq
This quantity can be further expanded at leading order in $\rs$ as
\be
\begin{split}
\label{Eq-EpsilonLargeRhoStar}
    \e & \simeq \frac{\sqrt{3}}{{64} \sqrt{2}} \, \frac{\mp^{2}}{T_3} \, \frac{ e^{4 \ruv/3} \, e^{- 8 \rs /3}}{ N_c c_{+}^{4} \sqrt{8 \ruv - 1}} \, \left( \frac{1}{\sinh(2\r)} \, \frac{\6 Q^2}{\6 \r} \right)^2 \\
 & = \frac{\sqrt{3}}{{16} \sqrt{2}} \, \frac{\mp^{2}}{T_3} \, \frac{N_{c}^{3} \, e^{4 \ruv/3} \, e^{- 8 \rs /3}}{c_{+}^{4} \sqrt{8 \ruv - 1}} \left( \frac{(2\r \coth(2\r)-1) (\sinh(4\r)-4\r)}{\sinh^{3}(2\r)} \right)^2 \, .
\end{split}
\ee
This expression exhibits a \emph{scaling behavior}: all the explicit dependence on $\rho$ is contained in the function inside the parenthesis, which does \emph{not} depend on the remaining parameters that instead all appear in the overall prefactor. The maximum of the function of $\rho$ inside the parenthesis occurs at $\r_{0} \simeq 0.98$. This is precisely the position at which the force acting on the brane has its maximum, corresponding to the inflection point for the brane potential --- see the discussion around Eq.~\eqref{Eq-InflectionPointLocation}. Notice that the value of $\epsilon$ is very suppressed by powers of $c_+$ and also has an exponential suppression in $\rs$: these small factors make the parameter $\epsilon$ very small even near the maximum of the force, around $\rho \sim \r_0$.

\smallskip

For the parameter $\eta$ we have, in the simultaneous  limit of large $c_+$ and $\rs$,
\be
\begin{split}
\label{Eq-EtaLargeRhoStar}
 \eta & \simeq \frac{\sqrt{3}}{8 \sqrt{2}} \, \frac{\mp^{2}}{T_3} \, \frac{e^{4 \ruv/3}}{N_c c_{+}^2 \sqrt{8 \ruv -1}} \, \left( \frac{1}{\sinh^{2}(2\r)} \, \frac{\partial^{2} \, Q^{2}}{\partial \r^{2}} + \frac{1}{\sinh(2 \r)} \, \frac{\partial \, (\sinh(2\r))^{-1}}{\partial \r} \, \frac{\partial \, Q^{2}}{\partial \r} \right) \\
 & = \frac{\sqrt{3}}{{4} \sqrt{2}} \, \frac{\mp^{2}}{T_3} \, \frac{N_c \, e^{4 \ruv/3}}{c_{+}^2 \sqrt{8 \ruv -1}} \\ & \times \left\{\frac{1}{\sinh^{6}(2 \rho)} \left[-5 + 40 \rho^2 + 4 (1 + 6 \rho^2) \cosh{4 \rho} +  \cosh{8 \rho} - \rho (30 \sinh{4 \rho} + \sinh{8 \rho})\right]\right\},
\end{split}
\ee
which also exhibits a \emph{scaling behavior}: The expression in brackets depends uniquely on $\rho$ and is multiplied by a coefficient depending on the parameters of the model. Eq.~\eqref{Eq-EtaLargeRhoStar} implies that also $\eta$ is suppressed by $c_+$, but with a smaller power than the suppression of $\epsilon$. The expression in brackets vanishes for $\r_0 \simeq 0.98$. However, it generically becomes large in the region $\rho < \r_0$, so in order to have slow-roll inflation we have to focus on the region $\r_0 < \rho \ll \rs$, in which the $\eta$ parameter is negative. The latter is a nice extra since observations to favour a red tilt of the power spectrum \cite{Komatsu:2010fb}.

\begin{figure}[ht!]
\begin{center}
\centerline{
\includegraphics[scale=0.8]{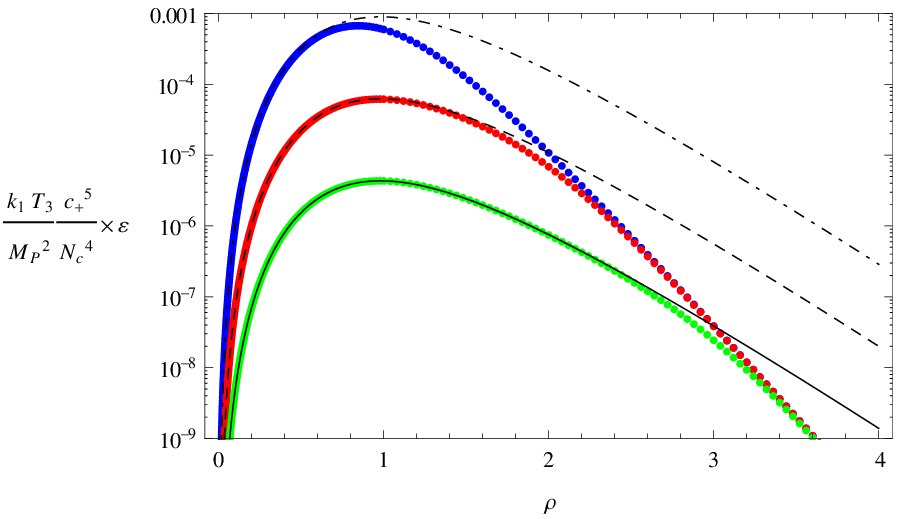}
\hspace*{0.4cm}
\includegraphics[scale=0.8]{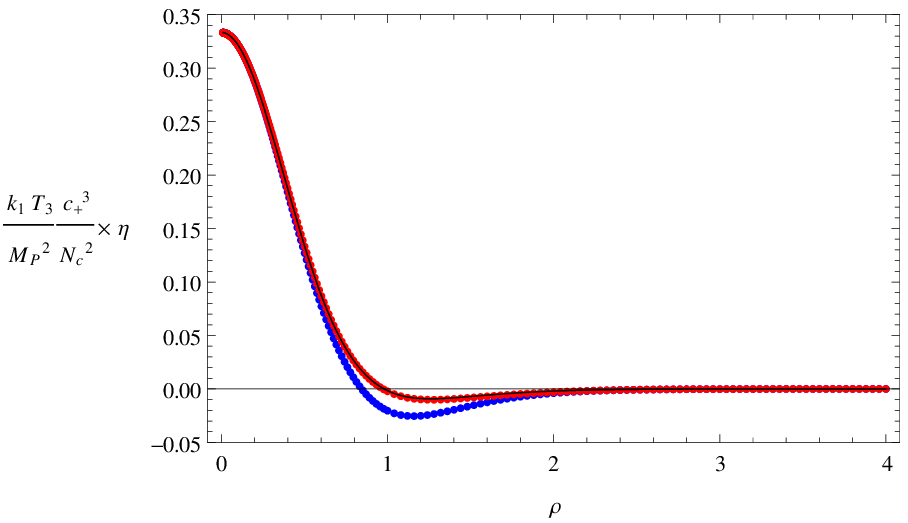}
}
\caption{Plots of $\e$ and $\eta$ rescaled by the indicated factors of $\frac{k_{1} T_{3}}{\mp^{2}} \frac{c_{+}^{5}}{N_{c}^{4}}$ and $\frac{k_{1} T_{3}}{\mp^{2}} \frac{c_{+}^{3}}{N_{c}^{2}}$, respectively, to remove the $c_{+}$-dependence. The curves with different colors correspond to solution obtained by numerically solving the master equation for three different values of $\rs \simeq$ $1$ (blue), $2$ (red) and $3$ (green). The solid, dashed and dotted black lines in the left figure display the approximation to $\e$ given in Eq.~\eqref{Eq-EpsilonLargeRhoStar}. In the right figure, we show the numerical solution for $\eta$ for $\rs \simeq$ $1$ (blue) and $2$ (red); the curve with $\rs \simeq 3$ would give the same result as the one with $\rs \simeq 2$. The solid black line shows the approximation for $\eta$ given in Eq.~\eqref{Eq-EtaLargeRhoStar}.}
\label{Fig-Walking-RescaledEpsilonEta}
\end{center}
\end{figure}

We show a comparison of the above approximate analytical expressions with numerical solutions to the master equation in Fig.~\ref{Fig-Walking-RescaledEpsilonEta}. In this way, one can verify the scaling behaviour and determine the quality of the above approximations. As one can see from Fig.~\ref{Fig-Walking-RescaledEpsilonEta}, the approximation for $\e$ is valid for $\r \lesssim \rs$, while the approximation for $\eta$ is a good one for all values of $\r$ already for $\rs \gtrsim 2$.

For $\rho \le \r_0$, our formulae for the inflaton potential and canonical normalization of the inflaton field are no longer a good approximation: the inflaton starts to move faster since the slow-roll approximation breaks down, and the DBI nature of the D3-brane action manifests.

Notice that while the profile for $\epsilon$ presents a single hump, the profile for $\eta$ in Fig. \ref{Fig-Walking-RescaledEpsilonEta}  has a  richer behavior; it decreases vanishing at $\rho_0\sim0.98$, then has a minimum at slightly larger value of the radial coordinate ($\rho\sim 1.26$), and then starts to increase towards a vanishing value for large $\rho$. Due to the aforementioned scaling behavior, one cannot change the radial profiles of the slow-roll parameters in the present treatment, but only their amplitude. It is nevertheless possible to add other sources to the geometry in the form of `flavour' D5-branes \cite{Conde:2011aa}, which can further modify these profiles (we will briefly comment on this possibility in Section~\ref{Sec-Discussion}).

\smallskip

\noindent

$\bullet$ {\bf Number of $e$-foldings:} The number of $e$-foldings, at leading order in a slow-roll expansion,  can be expressed as
\be
\label{Eq-NeWalking}
\begin{split}
N_{e} & = \int H dt \,\simeq\, \frac{1}{\mp^{2}} \int \frac{V}{\partial_{r} V}\,\left( \frac{d r }{d \rho}\right) d\rho = \frac{1}{\mp} \int \frac{1}{\sqrt{2 \e}} \left( \frac{d r}{d\r} \right) d\r \\
& \simeq { \frac{4 \sqrt{2}}{\sqrt{3}} } \, \frac{T_3}{\mp^{2}} \, \frac{c_{+}^{2} \,  \sqrt{8 \ruv - 1}\,e^{-4 \ruv/3}}{N_c} \int_{\rho_{0}}^{\rho_{inf}} \frac{\sinh^{4}(2 \r) \, d\r}{(2\r \coth(2\r)-1) (\sinh(4\r)-4\r)} 
\end{split}
\ee
where $\rho_{inf}$ (lying in the interval $\r_0 < \rho_{inf} \ll \rs$) can be adjusted to provide $60$ $e$-foldings of inflation.  Within this interval, we are ensured that our formula is correct since we are within the slow-roll regime. Note also that again there is a scaling behaviour --- the model parameters enter only into the overall prefactor.

\begin{figure}[h!]
  \begin{center}
    \includegraphics[width=7 cm, height= 4 cm]{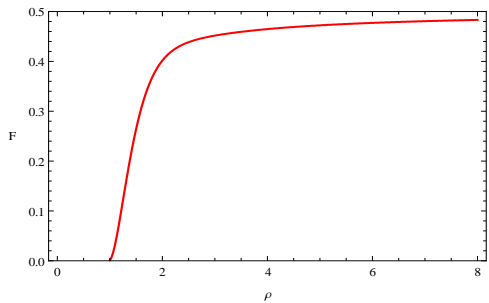}
  \end{center}
  \vspace{-0.6 cm}
  \caption{The function $F(\rho)$ appearing in Eq.~\eqref{Eq-RelationEtaNe}.}
  \label{Fig-Walking-F}
\end{figure}

Once properly normalized, $\rho_{inf}$ is associated to the position of the D3 brane $60$ $e$-folds before the end of slow-roll inflation. Comparing the expression for the $\eta$-parameter, Eq.~\eqref{Eq-EtaLargeRhoStar}, with the previous formula Eq.~\eqref{Eq-NeWalking}, we find a remarkable relation:
\be
\label{Eq-RelationEtaNe}
\lvert \eta \rvert  = \frac{1}{N_e} \, F(\rho_{inf}) \, ,
\ee
where $F(\rho_{inf})$ is a mild function of $\rho_{inf}$, whose values are between zero and (about) one half: see Fig.~\ref{Fig-Walking-F}. It is obtained by multiplying the absolute values of the two $\rho$-dependent functions in Eqs.~\eqref{Eq-EtaLargeRhoStar} and \eqref{Eq-NeWalking} since the prefactors depending on the parameters 
 in $\eta$ and $N_{e}$ are precisely the inverse of each other. Hence, for values of  $\rho_{inf}$ larger than say $3$, the parameter $\eta$ is $\sim - 1/(2 N_{e})$.

\smallskip

\noindent

$\bullet$ {\bf Fixing the normalization of the power-spectrum:} Besides realizing at least $60$ $e$-folds of inflation, we also have to ensure that the amplitude of the scalar power spectrum of primordial fluctuations produced during inflation matches the observed value \cite{Komatsu:2010fb}. This amounts to require that the following condition is satisfied 
\be
\left(\frac{V^{\frac{3}{2}}}{ V'\,\mp^3}\right) \, \Bigg\vert_{\r = \r_{inf}} = 2.4 \times 10^{-4} \quad \Rightarrow \quad \frac{V}{2\,\mp^4\, \epsilon} \, \Bigg\vert_{\r = \r_{inf}} = 5.6 \times 10^{-8} \, .
\ee
By tuning the available parameters of the model (for example the quantity $\ruv$) it is not hard to satisfy this condition. As we will  comment below, 
the tilt $n_{s}$ of the scalar power spectrum can also be easily accommodated to fit observations.

\smallskip

To conclude this section, let us emphasize once again that we have enough free parameters available for tuning the size of the slow-roll parameters independently. The geometrical features of the throats we consider are rich enough to allow for a flexible implementation of an inflationary scenario. The parameter $c_+$ controls the profile of the dilaton, $\rs$ is the position at which the $F_5$ and $H_{3}$ fluxes get turned on (see also \cite{Elander:2011mh}), while the parameter $\ruv$ controls the position at which our throat is glued to a compact space. In the QFT dual, the dynamics in the slow-roll inflationary region is mostly controlled by the VEVs of two operators, which are absent in the KS case: a dimension-2 VEV that brings the background on the baryonic branch \cite{Butti:2004pk} and the VEV for a dimension-6 operator specifically associated with the walking dynamics. Although we have freedom to choose these parameters, our scenario leads to robust predictions: the slow-roll parameter $\epsilon$ is much smaller than the $\eta$-parameter; the $\eta$-parameter is negative, and its absolute value is proportional to the inverse of the number of e-folds.

\subsection{An explicit realization of  an inflationary trajectory}

Let us apply these formulae  to determine a concrete example of an inflationary trajectory in the range $\rho_0 \le \rho \ll \rs$. We  focus on  $\rho\ll\rs$ since in this region our analytic expressions are accurate. When approaching $\rho \simeq \rs$ our approximations break down; in this region the geometry approaches very well the KS throat where the force on the brane vanishes. In the region $\rho<\rho_0$, on the contrary, the attractive force on the brane is typically too strong to support slow-roll inflation. The $\eta$-parameter becomes large and the inflationary evolution cannot gain many $e$-foldings in this region. We expect that once the brane enters the region $\rho < \rho_0$ it rapidly falls towards the tip of the throat (at $\r = 0$), where it starts backreacting on the geometry making inflation end (see Appendix~\ref{App-CommentsBackreaction} for a more quantitative treatment of this backreaction). Hence, we consider $\rho\sim \rho_0$ as the lower value of the radial coordinate at which standard slow-roll inflation ends.

Given these considerations, we would like to determine an inflationary trajectory generating sixty $e$-folds of slow-roll inflation nearby the inflection point for the potential, namely for $\rho \gtrsim \rho_0$. As we learned in the previous section, the slow-roll parameters are suppressed by powers of $\nicefrac{1}{c_+}$ and $e^{-\rs}$. In a regime of small values for these quantities, as the one we are interested in, it is not difficult to adjust the slow-roll parameters $\eta, \epsilon$ to sufficiently small values and moreover tune them independently, maintaining the robust relation $\epsilon\ll |\eta|$. The normalization of the power spectrum, then can be easily further tuned by choosing $\ruv$. (Note that the scalings of all quantities depend only on certain ratios of $c_{+}$ and $N_{c}$ such that any change in $N_{c}$ can be compensated by adjusting $c_{+}$ accordingly.)

\smallskip

Before providing the actual numbers, let us give an argument to fix the value of $T_3$, the tension of the moving D3 brane, which depends on the details of bulk moduli stabilization. Upon introducing a finite UV cutoff $\ruv$ and gluing the throat into a compact space, the graviton zero mode will have most of its support in the essentially unwarped compact space since it is exponentially suppressed in strongly-warped regions (see \eg the discussion around Eq. (C.10) of \cite{Kachru:2003sx}). That is, we can approximate the (four-dimensional Planck-mass) $\mp$ as
\be
  \label{Eq-CompactVolumePlanckMassRelation}
  \mp^{2} \approx \frac{2 \, \mcV_{6}}{\left(2 \pi \right)^{7} \, \alpha'^{4} \, \gs^{2}} \, ,
\ee
where $\mcV_{6}$ is the volume of the compact space. 

Recall that when defining the backgrounds around Eq.~\eqref{Eq-WrappedD5Metric} we have set $\ap \gs = 1$. This was used in deriving the above expressions for $N_{e}$, $\e$ and $\eta$, so  now we have to reinstate appropriate powers of $\ap \gs$ to make $N_{e}$, $\e$ and $\eta$ dimensionless. One can easily see that the relevant combination that should appear is
\be
  \frac{T_{3} \, \ap \gs}{\mp^{2}} \, ,
\ee
where the D$3$-brane tension $T_{3}$ is given by

\be
 T_{3} = \frac{1}{(2 \pi)^{3} \apsq \gs} \, ,
\ee
and $\mp$ is determined by Eq.~\eqref{Eq-CompactVolumePlanckMassRelation}. For the numerical estimates in the following, we assume the bulk moduli controlling $\gs$ and $\mcV_{6}$ have been stabilized such that $\gs \sim 0.1$ and $\mcV_{6} \sim \left(5 \, \sqrt{\ap} \right)^{6}$, which yields
\be
  \frac{T_{3} \, \ap \gs}{\mp^{2}} \approx \frac{(2 \pi)^{7} \a'^{4} \gs^{2}}{2 \, \mcV_{6}} \, \frac{\ap \gs}{(2 \pi)^{3} \apsq \gs} \sim \frac{(2 \pi)^{4}}{2} \, \gs^{2} \, \frac{\a'^{3}}{\mcV_{6}} \sim 5 \times 10^{-4} \, .
\ee
Similarly, the ratio $T_{3} / \mp^{4}$ is given by
\be
  \frac{T_{3}}{\mp^{4}} \approx \frac{\left( 2 \pi \right)^{11} \gs^{3} \, \a'^{6}}{4 \mcV_{6}^{2}} \sim 6 \times 10^{-4} \, .
\ee
Equipped with these two ratios (the latter one enters into the amplitude of the scalar power spectrum), whose precise values depend on the details of bulk moduli stabilization, we can now fix our parameters in order to match observations.

 \smallskip

\begin{figure}[ht!]
\begin{center}
\centerline{
\includegraphics[scale=0.6]{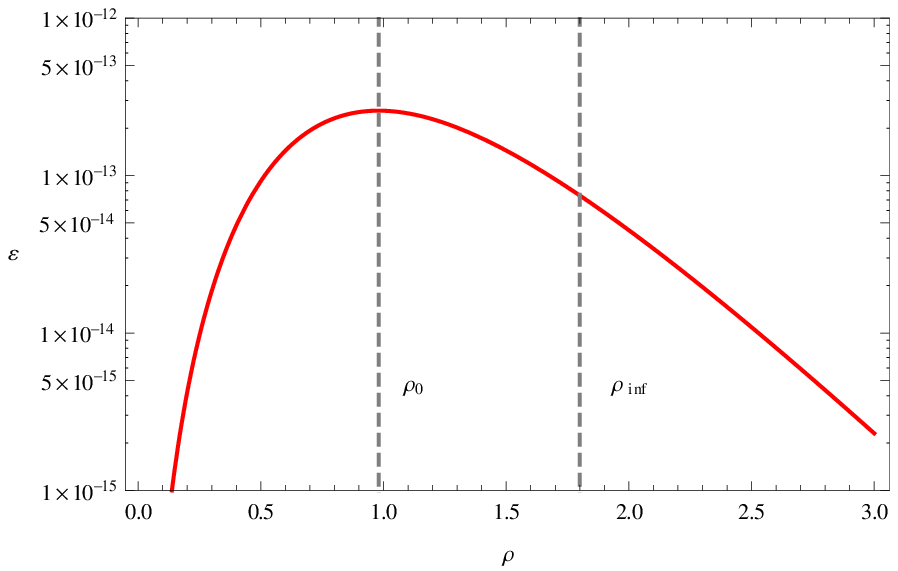}
\hspace*{0.4cm}
\includegraphics[scale=0.6]{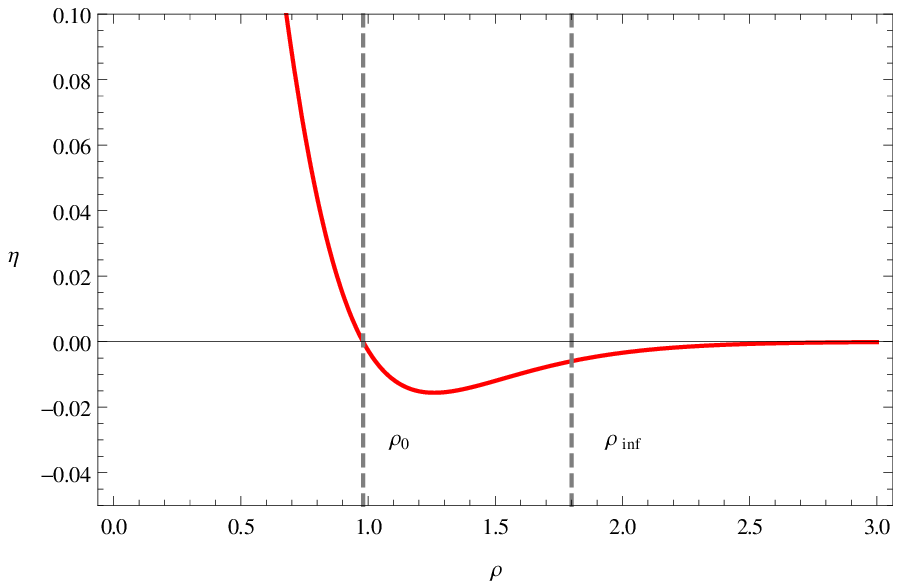}
\hspace*{0.4cm}
\includegraphics[scale=0.6]{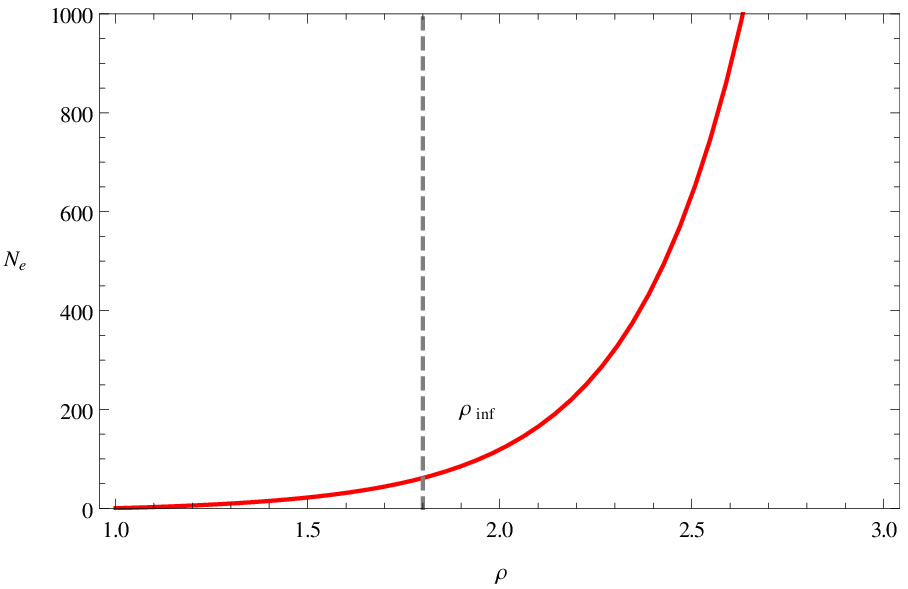}
}
\caption{Plots of $\e$, $\eta$ and $N_{e}$ for the example choice of parameters specified in Eq.~\eqref{Eq-ExampleParameters}. The gray dashed lines mark the positions of the inflection points $\r_{0} \simeq 0.98$ and the point $\r_{inf}$ corresponding to about $60$ $e$-folds before the end of inflation.}
\label{Fig-Walking-ExampleEpsilonEtaNe}
\end{center}
\end{figure}

We now choose the remaining parameters in order to realize inflation --- with mild tunings one easily finds a satisfactory inflationary trajectory. For example, the following choice works:
\begin{align}
\label{Eq-ExampleParameters}
  \rs & \simeq 2.90 \, , & \ruv & = 9 \, , & c_{+} & \simeq 6.1 \times 10^{3} \, , & N_{c} = 1 \, ,&\hskip1cm  \r_{inf}  \,= 1.8  
\end{align}
which yields $N_{e} \simeq 61.5$ and matches the observed power spectrum  normalization. At the value $  \r_{inf}  \,= 1.8  $, that is sixty $e$-folds before the end of slow-roll inflation, we find $\epsilon \sim 7 \times 10^{-14}$ and $\eta \sim -6 \times 10^{-3}$. Thus, the value for the scalar spectral index $n_{s} = 1 + 2 \eta - 6 \epsilon \simeq 0.988$ which is consistent with the value observed by WMAP within $2 \s$ \cite{Komatsu:2010fb}. The tensor-to-scalar ratio $r = 16 \, \e \sim 10^{-12}$ is completely negligible in accord with the non-observation of primordial gravitational waves; this is a generic feature of warped D3-brane inflation, cf. \eg \cite{Baumann:2006cd,Agarwal:2011wm,Dias:2012nf,McAllister:2012am}. See Fig.~\ref{Fig-Walking-ExampleEpsilonEtaNe} for plots of $\e$, $\eta$ and $N_{e}$ as a function of $\r_{inf}$ for the above choice of model parameters.

\smallskip

As we explained above, the $\eta$-parameter increases and becomes large for $\rho<\rho_0$; this implies that one does not gain many $e$-foldings while moving in this region. On the contrary, the $\epsilon$-parameter stays very small, and this would seem to imply that inflation cannot end in this set-up. However, as we discuss in Appendix~\ref{App-CommentsBackreaction}, when the brane reaches the deep IR, its backreaction on the geometry becomes important and our description of the throat is no longer reliable. In this regime deformations of the space-time should be so drastic to terminate inflation.


\section{Discussion}
\label{Sec-Discussion}

We have realized a model of slow-roll D-brane inflation in a warped
geometry dual to a field theory exhibiting walking behavior: hence its
name slow-walking inflation. In our set-up, the rich field content of
the  geometry we consider is associated with various parameters that
control an analytically calculable inflationary potential. It has the
correct properties to drive sixty $e$-folds of inflation and to match
the observed values for the amplitude and scale dependence of  the
power spectrum of scalar perturbations. Interestingly, the D-brane
potential generically has an inflection point, induced by the
particular shape of the dilaton profile. The relevant features of the
geometry have a counterpart in a dual field theory, allowing to
understand the characteristics of the inflationary brane potential in
terms of the combined action of operators with different
dimensions. Our model can be seen as a generalization of the
inflationary set-up in the  baryonic branch of the KS throat
\cite{Dymarsky:2005xt}: with that scenario, it shares the good feature of not needing an anti-D3-brane to drive inflation, since the warping and the fluxes turned on act with sufficient force on the brane to drive inflation. Moreover, inflation naturally occurs in a region that is easy to identify from the geometrical point of view, far away from the place in which the throat is glued to a compact space (as needed to couple gravity to the system and to add ingredients for stabilizing light moduli on the throat). Our methods can be straightforwardly applied to warped geometries that generalize the ones we considered here, containing \eg additional source D5-branes with non-trivial profiles in the throat 
 \cite{Conde:2011aa,Conde:2011rg,Nunez:2010sf}.
 This allows to further enrich the potential for the moving 
probe D3-brane, providing additional freedom to tune and engineer the inflationary process. It would also be very interesting to study dynamics of the brane along angular directions \cite{Easson:2007dh}, and analyze whether our set-up can be used to realize DBI inflation in the region in which the $\eta$-parameter becomes large \cite{Silverstein:2003hf}. We leave these questions for future work. 

\smallskip

Let us make some final cautionary remarks about the $\eta$-problem. The moving D3 brane couples with the background fields, hence it backreacts on the geometry. In Appendix \ref{App-CommentsBackreaction} we show that the backreaction on the throat fields is mild, until the brane reaches the far IR region. However, once the compact CY is glued to the throat, new couplings between bulk objects (introduced to stabilize light moduli) and the moving D3 brane itself, can modify the brane potential. All this can lead to order one corrections  to the inflationary $\eta$-parameter. The effects of gluing the compact CY to the throat can be seen also from the point of view of the dual field theory: new relevant operators become important in this case. For a complete model, it will be necessary to study the properties of these contributions. This is not easy to do and we leave it as an open problem for the future. Techniques as the ones developed in \cite{Baumann:2008kq,Baumann:2009qx,Baumann:2010sx} should be used for this purpose. Nonetheless, as we discussed above, our model has some new promising attractive features worth to study further. It exploits the rich geometrical content of the throat we considered to build a model of inflation, in which the brane potential is fully calculable. The resulting features of the inflationary potential can be tuned with much freedom thanks to the many parameters controlling the geometry.

Optimistically, this might imply that even after including the aforementioned corrections due to physics in the compact space, there is still room left for changing the parameters characterizing our original model in such a way that sufficient slow-roll inflation can  be achieved without qualitatively changing much the framework we developed. We leave these interesting questions for future work.

\acknowledgments

We thank Anupam Mazumdar and  Ivonne Zavala for useful discussions. GT
is supported by an STFC Advanced Fellowship ST/H005498/1. JE  and GT
thank the ESF HoloGrav Network for hospitality at Swansea University
during the Gauge/Gravity Duality conference there in April 2012, while
SH thanks both Swansea University and  ICG Portsmouth, for their warm
hospitality. JE and SH acknowledge funding through the German `Cluster
of  Excellence for Fundamental Physics: Origin and Structure of the
Universe'. SH also acknowledges support by and participation in the SFB-Transregio (TR33) `The Dark Universe'. CN is grateful for hospitality at CERN and to DAAD for an exchange fellowship to visit the MPI for Physics, Munich.


\appendix

\renewcommand{\theequation}{\thesection.\arabic{equation}}
\setcounter{equation}{0}

\section{Details of the backgrounds}
\label{App-BackgroundDetails}

In this Appendix, we give some details about the backgrounds sketched in Eqs.~\eqref{Eq-WrappedD5Metric} and \eqref{Eq-RotatedBackgroundMetricF5}. As explained in the main text, we consider the geometry produced by stacking on top of each other $N_c$ D5-branes that wrap an $S^2$ inside a CY-cone, \eg the conifold. Let us define a set of $SU(2)$ left-invariant one-forms as
\begin{align}
\label{Eq-SU2Inv1Forms}
\tilde{\w}_1 & = \cos\psi d\tilde\theta + \sin\psi\sin\tilde\theta d\tilde\phi \, , & \tilde{\w}_2 & = -\sin\psi d\tilde\theta +\cos\psi\sin\tilde\theta d\tilde\phi \, , & \tilde{\w}_3 & = d\psi + \cos\tilde\theta d\tilde\phi \, .
\end{align}
The conifold with topology $\mathbb{R} \times S^{2} \times S^{3}$ has a $SU(2) \times SU(2) \times U(1)$ isometry.

We assume that the functions appearing in the background depend only the radial coordinate $\r$ (the range of the angles is $0\leq \theta,\tilde{\theta}<\pi\,,\, 0\leq\phi,\tilde{\phi}<2\pi\,,\,0\leq\psi<4\pi$). We write the background (in Einstein frame) as,
\be
\label{Eq-WrappedD5Ansatz}
\begin{split}
ds^2 & =  e^{\Phi(\rho)/2} \left[ dx_{1,3}^2 + ds_6^2 \right] \, , \\
ds_6^2 & =  e^{2k(\rho)}d\rho^2 + e^{2 q(\rho)} (d\theta^2 + \sin^2\theta d\phi^2) \\
& \quad + \frac{e^{2 {g}(\rho)}}{4} \left[(\tilde{\omega}_1+a(\rho)d\theta)^2 + (\tilde{\omega}_2-a(\rho)\sin\theta d\phi)^2\right] + \frac{e^{2 k(\rho)}}{4} (\tilde{\omega}_3 + \cos\theta d\phi)^2 \, ,\\
F_{3} & = \frac{N_c}{4}\Bigg[-(\tilde{\omega}_1+b(\rho) d\theta)\wedge (\tilde{\omega}_2-b(\rho) \sin\theta d\phi)\wedge (\tilde{\omega}_3 + \cos\theta d\phi) \\
& \quad + \partial_\rho b \ d\rho \wedge (-d\theta \wedge \tilde{\omega}_1  + \sin\theta d\phi \wedge \tilde{\omega}_2) + (1-b(\rho)^2) \sin\theta d\theta\wedge d\phi \wedge \tilde{\omega}_3\Bigg] \, .
\end{split}
\ee
The full solution is determined by solving the equations of motion for the functions $\{ a, b, \Phi, g, q, k \}$. As described in the main text, the BPS equations derived using this ansatz can be rearranged in a more convenient form, by rewriting the functions appearing in Eq.~\eqref{Eq-WrappedD5Ansatz} in terms of a new basis of functions where the BPS equations decouple, as explained in \cite{HoyosBadajoz:2008fw}, in which the functions $\{ a, b, g, q, k \}$ read
\be
\label{Eq-NewFunctions1Appendix}
\begin{split}
4 e^{2q} & = \frac{P^2-Q^2}{P\coth(2\r) -Q} \, , \quad e^{2{g}} = P\coth(2\r) -Q \, , \\
e^{2k} & = \frac{P'}{2} \, , \quad a = \frac{P}{\sinh(2\rho)\left(P\coth(2\r) -Q\right)} \, , \quad b = \frac{2 \rho}{\sinh(2 \rho)} \, .
\end{split}
\ee
Using these new variables, one can work with the BPS equations to obtain a single second order equation for $P(\rho)$, while the quantities $Q$ and $\Phi$ are given by
\be
\begin{split}
\label{Eq-NewFunctions2Appendix}
Q(\rho) & = N_c (2\rho \coth (2\rho) -1) \, , \\
e^{4\Phi} & = \frac{8\,e^{4\Phi_0} \,\sinh^2(2\rho)}{(P^2-Q^2) P'} \, ,
\end{split}
\ee
with $\Phi_0$ a constant and $P$ determined by the second order equation mentioned above:
\beq
P'' + P' \left( \frac{P'+Q'}{P-Q} +\frac{P'-Q'}{P+Q} - 4 \coth(2\rho) \right) = 0 \, .
\label{Eq-MasterEquationAppendix}
\eeq
In the case of the background after the rotation, the details go as follows. We use the vielbein in Einstein frame given by Eq.~\eqref{Eq-Vielbein}. The full generated configuration is obtained as
\be
\label{Eq-RotatedBackground}
\begin{split}
ds_{E}^2 & = \sum_{i=1}^{10} (e^i)^2 \, , \\
F_3 & = \frac{e^{-\frac{3\Phi}{4}}}{\hat{h}^{\frac{3}{4}}} \left[f_1 e^{123} + f_2 e^{\theta\varphi 3} -f_3(e^{\varphi 1 3} + e^{\theta 23})+ f_4(e^{\r 1 \theta} + e^{\r \varphi 2}) \right] \, , \\
B_2 & = k_2 \frac{e^{3\Phi/2}}{\hat{h}^{1/2}} \left[ e^{\r 3}+ \cos\mu \, (e^{\theta\varphi} +e^{12}) +\sin\mu \, (e^{\varphi 1} +e^{\theta 2}) \right] \, , \\
H_3 & = - k_2 \frac{e^{\frac{5\Phi}{4}}}{\hat{h}^{\frac{3}{4}}} \left[ -f_1 e^{\theta\varphi \rho} - f_2 e^{12\r}+ f_3(e^{\theta 2 \r} + e^{\varphi 1 \r}) - f_4(e^{\theta 1 3} - e^{\varphi 2 3}) \right] \, , \\
F_5 & = k_2 \frac{d}{d\r} \left(\frac{e^{2\Phi}}{\hat{h}} \right) \hat{h}^{3/4} e^{-k-\frac{5 \Phi}{4}} \left[-e^{tx1x2x3 \r}+ e^{\theta\varphi 1 2 3} \right] \, ,
\end{split}
\ee
where
\be
  \cos\mu = -\frac{P- Q\coth(2\r)}{P\coth(2\r) -Q} \, ,
\ee
the functions $f_i, i=1,..,4$ are given by
\be
\begin{split}
f_1 & = -2 N_c e^{-k-2g} \, , \quad f_2 =\frac{N_c}{2}(a^2-2ab +1)e^{-k-2q} \, , \\
f_3 & = N_c(b-a)e^{-k-q-g} \, , \quad f_4=\frac{N_c}{2}b' e^{-k-q-g} \, ,
\end{split}
\ee
and we denoted
\beq
e^{i j k \dots l} = e^{i} \wedge e^{j} \wedge e^{k} \wedge \dots \wedge e^{l} \, .
\eeq
A necessary condition to apply this solution-generating technique is that the quantity $e^{\Phi}$ is bounded from above, being an increasing function with $e^{\Phi(\infty)}$ its maximum value. This condition can be linked with the absence of D7-brane sources in the configuration of Eq.~\eqref{Eq-RotatedBackground} --- see \cite{Gaillard:2010qg} for details.

\section{Some comments on the dual QFT \& brane backreaction}

Here we collect some arguments that should be thought as `plausibility arguments' for the results obtained in this paper. The first of these arguments, which is of qualitative nature, attempts a QFT explanation for the form of the potential and the force felt by the travelling D3-brane. The second argument aims at justifying that our set-up, with an inflection point around which inflation starts, does not get substantially modified when including in the effects of back reaction of the travelling D3-brane on the geometry.

\subsection{A field-theoretical view on the form of the potential and the force}
\label{App-CommentsDualQFT}

The goal of this Appendix is to get some handle on the force acting on the moving D3 brane using arguments in the dual QFT. The argument we will present is of qualitative nature. We will basically only attempt to make the point that the `shape' of the force on the D3 as computed in the gravitational set-up has some connection with the same quantity calculated in a QFT that \emph{approximately} describes the process.

In order to make this discussion more self-contained, we will summarise some aspects of the minimally supersymmetric QFT associated with the rotation of the walking solution as introducted in Section~\ref{Sec-GeneratingRotationUDuality}.

Let us start by considering the quiver field theory
\beq
SU((k+1)N_c + n_f) \times SU(k N_c + n_f) \, ,
\label{Eq-Quiver1}
\eeq
where in this case $n_f=1$ represents the single D3-brane that is moving along the throat, representing our candidate inflaton field. 

For the case $n_f=1$, the field theory was well-studied by Seiberg in the 1990's (see for instance \cite{Intriligator:1995au}). In our case, we consider a supersymmetric QFT with quiver
\beq
SU((k+1)N_c) \times SU(k N_c) \, ,
\label{Eq-Quiver2}
\eeq
and add one \emph{probe} D3-brane. This breaks SUSY generating a non-constant potential and a force on the probe D3.

This type of SUSY breaking may be thought of as being `soft'. We will therefore use Seiberg's supersymmetric formalism, knowing that the modification due to soft breaking will not change the qualitative results --- in other words, in our model, we assume that the scale of SUSY breaking is smaller than the strong coupling scale of the QFT in Eq.~\eqref{Eq-Quiver2}. This was done successfully in this context in the papers \cite{Aharony:1995zh}.

Let us go to the point now and summarise some things about the QFT's described above, adapting arguments of \cite{Dymarsky:2005xt}. The correct indices in the following expressions are explicitly written in \cite{Dymarsky:2005xt}.

\begin{itemize}
\item The field theory of Eq.~\eqref{Eq-Quiver1} has a superpotential given by a tree-level part and a non-perturbative part of the form
\beq
W \sim M^2 + \frac{1}{\Lambda} \left( M \mcA\mcB - \det[M] \right) \, ,
\label{Eq-Superpotential}
\eeq
where the first term corresponds to the tree-level superpotential 
\beq
W_{tree} \sim ABAB \sim MM \, .
\eeq
Here, $A, B$ are the usual bifundamentals joining the two gauge groups and $M$ is the meson superfield dual to the position of the D3-brane probe \cite{Dymarsky:2005xt}. $\mcA, \mcB$ are the baryon and anti-baryon fields, that in the case of the quiver in Eq.~\eqref{Eq-Quiver2} are given by 
\begin{equation}
\label{Eq-BaryonicOperators}
\mcA = \epsilon^{i_1 \dots i_N} A_{i_1} \dots A_{i_N} \,, \quad \mcB = \epsilon^{i_1 \dots i_N} B_{i_1} \dots B_{i_N} \, .
\end{equation}
For the quiver in Eq.~\eqref{Eq-Quiver2}, $N = 2 N_c$ at the end of the cascade where the quiver becomes $SU(2N_c) \times SU(N_c)$ with the $SU(2N_c) $ factor being strongly-coupled. In the case of the quiver in Eq.~\eqref{Eq-Quiver1}, the last step of the cascade gives the gauge group $SU(2N_c + n_f) \times SU(N_c + n_f)$. In the strongly-coupled sector, the number of colors is $2N_c + n_f$, while the number of flavors is $2N_c + 2 n_f$. Since the index $N$ in Eq.~\eqref{Eq-BaryonicOperators} runs over flavor indices, then for $n_f =1$ (one probe D3-brane), the baryon operators have one free flavor index.

\item For simplicity, we assume that the K\"ahler potential is trivial. This  implies that up to numerical factors, the potential reads 
\beq
\label{Eq-VAppendix}
V\sim \left( \frac{\partial W}{\partial M} \right)^2 + \left( \frac{\partial W}{\partial \mcA} \right)^2 + \left( \frac{\partial W}{\partial \mcB} \right)^2 .
\eeq

\item We will also consider that the difference between the VEVs of the baryon and anti-baryon fields is related to the VEV of the operator $\mcU$, defined in \cite{Dymarsky:2005xt},
\beq
\mcU \sim ( \mcA-\mcB ) \, .
\eeq
This is the case for example in the KS-limit of the quiver, where $\mcU = 0$.

\item A crucial point is that the operator $\mcU$ is related to an invariant of the rotation procedure described in Section~\ref{Sec-GeneratingRotationUDuality}. This invariant is the function
\beq
\label{Eq-FAppendix}
F = a^2+ 4 e^{2h-2g}-1 \, ,
\eeq
where the functions $a(\r), h(\r), g(\r)$ are defined in Eq.~\eqref{Eq-NewFunctions1Appendix}. The relation between $\mcU$ and this particular combination of gravity fields above is strengthened by the fact that when analysing the asymptotics of the field $F$ above, it is found that it describes a dimension-2 operator getting a VEV \cite{Elander:2011mh}. The function $F$, an invariant under the rotation procedure, is drawn in Fig.~3 of \cite{Elander:2011mh}.

\item We also assume that the meson matrix has one non-zero component while the remaining components vanish, such that $\det M=0$.

\end{itemize}

Under these assumptions, we now calculate the potential from Eq.~\eqref{Eq-VAppendix} and obtain
\beq
V \sim ( M-\frac{1}{\Lambda} \mcA \mcB )^2+ M^2(\mcA^2+{\cal  B}^2) \, .
\eeq
From this we may calculate the force on the meson field $f_M \sim - \partial_M V$. We consider the situation in which $\langle M \rangle$ changes slowly on a background of fixed
 $\langle \mcA \rangle, \langle \mcB \rangle$, we then have
\beq
f_M \sim \mcA^2 + \mcB^2 - \mcA \mcB \sim (\mcA - \mcB)^2 \sim \mcU^2
\eeq
Let us summarise: we are proposing a \emph{very qualitative} reasoning to get a handle on one of the scales of the problem and its effect on the force on the probe D3. From the QFT perspective, we are dealing with a case of very soft SUSY breaking. For a field theory like the one in Eq.~\eqref{Eq-Quiver1}, for $n_f=1$, we have a potential like in Eq.~\eqref{Eq-Superpotential}. But we are studying this by starting from a theory like in Eq.~\eqref{Eq-Quiver2} and using the fact that this theory has VEV's for the baryon and anti-baryon superfields. Then we break SUSY `softly' by the addition of a \emph{moving} D3 brane. Hence, we propose that we have `background' baryonic fields taking a large VEV,
while the mesonic superfield probes this QFT, varying slowly. The result of this very qualitative reasoning tells us that the force on the D3 probe has a similar shape as that of the `geometry' invariant given by the combination in Eq.~\eqref{Eq-FAppendix}. In terms of the radial coordinate, this is a function which has a maximum close to an intermediate point --- that turns out to be $\rs$ --- and then decays in the IR and in the far UV. Notice that the QFT dual to our geometry has, as discussed in \cite{Elander:2011mh}, two \emph{independent scales}. One of the scales controlled by $\rs$ indicates, when flowing from the UV to the IR, the beginning of the walking region. Flowing further towards the IR, the second scale is given by $\r_I \sim 1$. This scale indicates the end of the walking region, the onset of confinement and $\mathbb{Z}_{2 N_c}$ symmetry breaking. These two scales, not surprisingly control our inflationary dynamics.

Note that we obtain a qualitatively very similar result for the $\epsilon$ slow-rolling parameter in Eq.~\eqref{Eq-SlowRollParametersRho}. This is due to the fact that roughly $\epsilon \sim (V'/V)^2$, for an almost constant $V$. In this sense, there should be some connection between the QFT potential and the potential felt by the D3 probe. Note that the point in the radial direction where the force is strongest corresponds to the transition between the QFT being dominated by its UV behaviour to it being dominated by its IR behaviour. This transition point dominates the behaviour of our model of inflation.

Let us move now to a more quantitative and precise way of understanding the effects of the backreaction of the D3 brane on the throat.

\subsection{Backreaction of the travelling D3-brane}
\label{App-CommentsBackreaction}

The aim of this subsection is to explore how good the approximation of a travelling \emph{probe} D3-brane, actually is. The outcome of the study is that, for slowly moving D3 branes, the backreaction effects are important only close to the far IR region $\rho \sim 0$. In other words, we will give a quantitative way of calculating the backreaction of the D3-brane on the geometry, in a particular situation --- this will be in concordance with the approximations we used in our model of inflation.

To make this calculation feasible, we assume for simplicity to be in a KS-background (since we are taking very large values of $c_+$ this  is not a very drastic assumption) and add D3 sources. We will also assume that the D3-brane moves quite slowly. That is, its kinetic energy is negligible when calculating the deformation of the space time that is mostly sourced by its mass.

The main simplification is to consider the  backreaction assuming the D3 to be smeared on the directions $\theta,\varphi, \tilde{\theta},\tilde{\varphi},\psi$. If we  focus on this particular and simplified problem, we can write an \emph{exact} solution. The solution is actually written in Eqs.~(3.8) and (3.9) of \cite{Conde:2011aa}. There, it was found that the warp factor $\hat{h}$ becomes
\beq
\hat{h}=\hat{h}_{KS} + \frac{4 k_1}{\epsilon^{8/3}} \int_\rho^\infty \frac{S(x)}{[\sinh(4x)-4x]^{1/3}},\;\;\;\;\; \hat{h}_{KS}=
\frac{2^{5/3} k_1 N_c^2}{\epsilon^{8/3}}\int_{\r}^{\infty}dx \frac{(\sinh(4x)-4x)^{1/3} (2x \coth(2x)-1)}{\sinh^2(2x)}, 
\eeq
where $S(x)$ is the probe D3 distribution. We should take $S(\r) = \delta(\r - \r_0)$ or some other function localized close to the position of the probe D3, $\rho_0$, say like a gaussian or similar. The point $\rho_0$ (which should not be confused 
with the position of the inflection point introduced in Section~\ref{Sec-ScalingBehaviour}) {\it slowly }
changes as the brane moves towards the end of the geometry.

Suppose that we want to estimate the backreaction for large values of the radial coordinate. In this case the warp factor will change as
\beq
\hat{h} - \hat{h}_{KS} \propto \frac{1}{[\sinh(4\rho_0)-4\rho_0]^{1/3}} \sim \frac{1}{e^{4\rho_0/3}}
\label{Eq-BackreactedWarpFactorUV}
\eeq
where we approximated the profile of branes by a delta function localized at $\rho_0$. Thus, the correction is very small. However, for the case in which the brane is close to $\rho \sim 0$, we have that the new contribution to the warp factor goes like
\beq
\hat{h} - \hat{h}_{KS} \propto \frac{1}{[\sinh(4\rho_0) - 4\rho_0]^{1/3}} \sim \frac{1}{\rho_0}
\label{Eq-BackreactedWarpFactorIR}
\eeq
for small $\rho_0$. Hence the correction is large.

Now, to the point: for values where inflation occurs, we are able to use the approximation above and see explicitly that the D3 probe should actually not change things too much. After the end of Inflation, we come close to $\r = 0$ and the set-up and the solution should change due to the backreaction.

Things should not be very different for large but finite values of $c_+$, in which case an exact solution cannot be written, but the system is close to KS.

There is a field theoretical way of stating the argument above: if the meson field is `slowly varying', that is the D3 is slowly rolling, we can integrate out the meson in the super potential of Eq.~\eqref{Eq-Superpotential} and we would then get the super potential that one typically has in a quiver of the form in Eq.~\eqref{Eq-Quiver2}, giving the usual baryonic branch as one of its vacua. The gravity solution is not much different to the (baryonic branch of the) Klebanov-Strassler background, this is expressed by Eq.~\eqref{Eq-BackreactedWarpFactorUV}. As the meson VEV becomes comparable to the other scales in the problem, the meson field becomes more influential and we should instead describe the SUSY dynamics of the quiver in Eq.~\eqref{Eq-Quiver1}. This coincides with the D3-brane going to the far IR of the geometry and changing it considerably --- this is expressed by Eq.~\eqref{Eq-BackreactedWarpFactorIR}. In all the equations above, we were assuming a SUSY situation (or very `softly' broken SUSY as in \cite{Aharony:1995zh}).

\providecommand{\href}[2]{#2}\begingroup\raggedright

\end{document}